\documentclass[aps, prd, preprint, groupedaddress, nofootinbib]{revtex4-1}

\usepackage{graphicx}
\usepackage{epsfig}
\usepackage{bm}
\usepackage{amsfonts}
\usepackage{amsmath}
\usepackage{amssymb}
\usepackage{multirow}
\usepackage{subfigure}
\usepackage{wasysym}
\usepackage{array}
\usepackage{diagbox}
\usepackage{makecell}
\usepackage{slashbox}

\begin{document}



\title{Inflation Based on the Tsallis Entropy}





\author{
Zeinab Teimoori$^{1}$\footnote{zteimoori16@gmail.com},
Kazem Rezazadeh$^{2}$\footnote{kazem.rezazadeh@ipm.ir},
and Abasat Rostami$^{3}$\footnote{aba-rostami@ipm.ir}
}

\affiliation{
$^{1}$\small{Department of Physics, University of Kurdistan, Pasdaran Street, P.O. Box 66177-15175, Sanandaj, Iran}\\
$^{2}$\small{School of Physics, Institute for Research in Fundamental Sciences (IPM), P.O. Box 19395-5531, Tehran, Iran}\\
$^{3}$\small{Physics and Energy Engineering Department, Amirkabir University of Technology, P.O. Box 159163-4311, Tehran, Iran}
}




\begin{abstract}

We study the inflationary scenario in the Tsallis entropy-based cosmology. The Friedmann equations in this setup can be derived by using the first law of thermodynamics. To derive the relations of the power spectra of the scalar and tensor perturbations in this setup, we reconstruct an $f(R)$ gravity model which is thermodynamically equivalent to our model in the slow-roll approximation. In this way, we find the inflationary observables, including the scalar spectral index and the tensor-to-scalar ratio in our scenario. Then, we investigate two different potentials in our scenario, including the quadratic potential and the potential associated with the natural inflation in which the inflaton is an axion or a pseudo-Nambu-Goldstone boson. We examine their observational viability in light of the Planck 2018 CMB data. We show that although the results of these potentials are in tension with the observations in the standard inflationary setting, their consistency with the observations can be significantly improved within the setup of the Tsallis entropy-based inflation. Moreover, we place constraints on the parameters of the considered inflationary models by using the current observational data.

\end{abstract}



\maketitle



\section{Introduction}
\label{section:introduction}

The first inflationary model was proposed by Starobinsky \cite{Starobinsky:1980te} in 1980 and it was based on the addition of the $R^2$ term in the Einstein-Hilbert action with the motivation to include semi-classical quantum effects to the gravity theory. It is interesting to point out that although this model is the first inflationary model, it is in very good agreement with the current observational data. Then, in 1981, Sato \cite{Sato:1980yn, Sato:1981ds}, suggested a scenario in which the Universe has undergone a rapid acceleration in the early stages of its evolution, and afterward it turned into a fireball with a very high temperature. Subsequently, in the same year, Guth \cite{Guth:1980zm} has shown that by including an accelerating phase before the radiation-dominated era, one can resolve the hot Big Bang cosmology problems such as the flatness problem, the horizon problem, and the magnetic monopole problem. Guth's model \cite{Guth:1980zm} is known as the old inflation and it is based on a scalar field that goes from a false vacuum towards a true vacuum through a first-order phase transition in the form of a quantum-tunneling process. The old inflation \cite{Guth:1980zm} suffers from some fundamental problems, and to resolve its problems, other inflationary models were suggested later \cite{Linde:1981mu, Albrecht:1982wi, Linde:1983gd, Linde:1986fd, Linde:1986fc}. For a nice review of the history of the first 30+ years of inflation, see \cite{Sato:2015dga}.

Along with the evolution of the inflationary models, the theory of cosmological perturbations was developed too \cite{Starobinsky:1979ty, Mukhanov:1981xt, Hawking:1982cz, Starobinsky:1982ee, Guth:1982ec, Starobinsky:1983zz}. The calculation of the primordial tensor perturbations in the early de Sitter stage was first elaborated by Starobinsky \cite{Starobinsky:1979ty} in 1979, in terms of the related quantity which is the spectrum of primordial GW background after the Hubble radius crossing at the radiation-dominated stage. His original motivation was physically sound with the aim to investigate the initial state of the Universe. Then, Mukhanov and Chibisov \cite{Mukhanov:1981xt} calculated the spectrum of the perturbations in the range of observable scales in the context of the Starobinsky $R^2$ inflation \cite{Starobinsky:1980te}. In addition, the quantitatively correct expression for the GWs spectrum produced in the Starobinsky model \cite{Starobinsky:1980te} was first presented in \cite{Starobinsky:1983zz}. The quantum fluctuations of the scalar field during inflation lead to the generation of the perturbations whose growth can seed the Large-Scale Structure (LSS) formation and the anisotropies observed in the Cosmic Microwave Background (CMB) radiation \cite{Starobinsky:1979ty, Mukhanov:1981xt, Hawking:1982cz, Starobinsky:1982ee, Guth:1982ec, Starobinsky:1983zz}. Thus, from the observations corresponding to the LSS formation and CMB anisotropies, we can obtain valuable information about the physics of the early Universe. Important observational results about inflation have been presented by Planck collaboration \cite{Planck:2018jri}, which are obtained from measurements of the CMB anisotropies in both temperature and polarization. Using these observational results, one can discriminate between different inflationary models.

The simplest inflationary model is based on a single scalar field minimally coupled to gravity \cite{liddle2000cosmological, lemoine2007inflationary}. The scalar field responsible for inflation is known as the inflaton. If the potential energy of the inflaton dominates over its kinetic energy, then the inflaton rolls downward its potential slowly. In the chaotic inflationary scenarios \cite{Linde:1983gd}, the slow-roll motion of the inflaton is provided by the Hubble friction term in the equation of motion of the scalar field, which is predominant in the early stages of the inflaton, and it then suppresses as the inflation goes towards its end. In the new inflation models \cite{Linde:1981mu}, however, the slow-roll phase of inflation is provided by a plateau-like potential for the inflaton field.

Inflation has occurred in the GUT energy scale which is about two orders of magnitude less than the Planck energy scale. Since inflation has occurred in the regime of high-energy physics, therefore it is expected that quantum gravitational effects had a decisive role in the dynamics of the early Universe. In those energy scales, the gravity theory may be modified due to the quantum gravitational effects. One way to determine the corrections to the gravity theory is the use of the gravity-thermodynamics conjecture implying that there is a deep connection between gravity and thermodynamics \cite{Jacobson:1995ab, Frolov:2002va, Padmanabhan:2002sha, Padmanabhan:2003gd, Eling:2006aw, Paranjape:2006ca, Kothawala:2007em, Padmanabhan:2009vy, Verlinde:2010hp}. This conjecture implies that using the thermodynamics laws for the Universe as a thermodynamical system, one can derive the gravitational equations governing the evolution of the Universe. In particular, the Friedmann equation can be derived from the first law of thermodynamics \cite{cai2005first, cai2008corrected, Sheykhi:2010zz, Sheykhi:2010yq, Sheykhi:2018dpn, Sheykhi:2021fwh}. If the entropy is assumed as the Hawking-Bekenstein entropy \cite{bekenstein1973black}, one can derive the standard Friedmann equation. If we consider some modifications to the entropy of the system, then the gravitational equations will be modified accordingly. In the early Universe, the entropy is expected to be different from the standard Bekenstein-Hawking entropy \cite{bekenstein1973black} due to the quantum gravitational effects that appear in the regime of high-energy physics.

One form that can be considered for the entropy of the Universe in its primordial stages is the Tsallis entropy $S_h \propto A^\beta$ \cite{tsallis2013black} in which $A$ is the horizon area and $\beta$ is a constant parameter known as the Tsallis parameter. This entropy is a generalization of Boltzmann-Gibbs entropy \cite{tsallis2013black, tsallis2019black}, and it has been suggested to solve a thermodynamic puzzle. The Boltzmann–Gibbs statistics is not capable of describing the systems with divergent partition functions such as the gravitational systems and requires a non-additive generalization of the entropy definition \cite{tsallis1988possible, lyra1998nonextensivity, tsallis1998role}. Accordingly, Tsallis and Cirto \cite{tsallis2013black, tsallis2019black} introduced an entropy expression leading to non-extensive statistics. The Tsallis entropy has attracted a high level of research interest over the years, and so far, remarkable results have been found based on this entropy in a lot of complex systems such as self-gravitating stellar systems \cite{plastino1993stellar, hamity1996generalized}, black holes \cite{tsallis2013black, tsallis2019black}, background radiation \cite{tsallis1995generalization}, neutrinos \cite{kaniadakis1996generalized, luciano2021q}, holographic dark energy \cite{tavayef2018tsallis, saridakis2018holographic, Nojiri:2021iko, Nojiri:2022aof, pandey2022new} and dark matter \cite{guha2019model}, thermodynamic gravity \cite{Sheykhi:2018dpn, geng2020modified, luciano2022baryogenesis}, low-dimensional dissipative systems \cite{lyra1998nonextensivity}, and polymer chains \cite{jizba2017tsallis}. In the special case where the Tsallis parameter is taken as $\beta=1+\Delta/2$ with $0\leq\Delta\leq1$, the Tsallis entropy reduces to the Barrow entropy \cite{Barrow:2020tzx}. The cosmological implications of the Barrow entropy have been regarded in the literature extensively in recent years (see, e.g., \cite{Anagnostopoulos:2020ctz, Barrow:2020kug, Dabrowski:2020atl, Saridakis:2020cqq, Jusufi:2021fek, Leon:2021wyx, Nojiri:2021jxf, Luciano:2022pzg, Luciano:2022ffn, Luciano:2022hhy, Sheykhi:2022gzb, Luciano:2023roh}).

In this paper, we investigate the implications of the Tsallis entropy for the inflationary phase of the early Universe. To provide the accelerated expansion of the Universe in the inflationary phase, we assume the matter-energy content of the Universe follows the form of a canonical scalar field that plays the role of the inflaton. We take the inflaton potential to be in the form of a simple quadratic potential $V(\phi)=m^2\phi^2/2$ which leads to a chaotic inflationary scenario \cite{Linde:1983gd}. In the framework of standard inflation, this potential is not favored in light of the current CMB data provided by the Planck 2018 collaboration \cite{Planck:2018jri}. We show that the results of this potential for the inflationary observables can be improved significantly in the context of Tsallis entropy-based inflation.

One another elegant inflationary scenario that provides a sensible mechanism to generate a flat potential is natural inflation \cite{Freese:1990rb}. In this class of models, the scalar field $\phi$ enjoys a shift symmetry $\phi \rightarrow \phi + {\rm const.}$, which is slightly broken explicitly or due to the non-perturbative quantum effects to a discrete symmetry $\phi \rightarrow \phi + 2\pi$ \cite{Freese:1990rb}. This feature gives rise to a periodic potential, as appropriate for inflation \cite{Freese:1990rb, svrcek2006axions}. Note that the presence of this symmetry protects the potential from radiative corrections. The scalar field with a flat potential originating from the shift symmetry is known as an axion. In this regard, the inflaton in the natural inflation is an axion or a pseudo-Nambu-Goldstone boson. The original natural inflation is described by a cosine-type periodic potential in the framework of a standard inflationary scenario based on the Einstein gravity where the entropy of the horizon obeys the Bekenstein-Hawking area law \cite{bekenstein1973black}. In this setup, the natural inflation is not very favored by the latest observation of the Planck 2018 collaboration \cite{Planck:2018jri}, and its results can satisfy only the 95\% CL constraint of Planck 2018 data \cite{Planck:2018jri}. This point motivates us to study the natural inflation in the setup of the Tsallis inflation and compare its predictions with the Planck 2018 constraints \cite{Planck:2018jri}.

Although the Friedmann equations of the Friedmann-Robertson-Walker (FRW) Universe in the framework based on the Tsallis entropy have already been derived in \cite{Sheykhi:2018dpn}, the study of inflation in this context also requires the equations of the power spectra of the primordial scalar and tensor perturbations. To this aim, we need the action of the gravity theory. In the absence of a unique action for our Tsallis inflation model, we construct an $f(R)$ gravity model which is thermodynamically equivalent to the Tsallis gravity in the slow-roll regime of inflation. The modified $f(R)$ gravity is a conceivable generalization of general relativity in which $f(R)$ is an arbitrary function of the Ricci scalar $R$. Over the years, the $f(R)$ gravity models have been extensively studied in the literature (see, e.g., \cite{Hwang:2001pu, Sotiriou:2008rp, DeFelice:2010aj, kaneda2010slow, Nojiri:2010wj, huang2014polynomial, Nojiri:2017ncd, cheong2020beyond, Nojiri:2020wmh, aziz2021inflationary, parrilla2023chameleon}). One of the most important motivations that led us to this choice is that it is possible in the $f(R)$ gravity to construct a model whose entropy is proportional to the powers of the horizon area, like the Tsallis entropy. Moreover, the $f(R)$ gravity models have phenomenological effective features that can describe inflation in the regime of high-energy physics. On one side $f(R)$ actions are general enough to cover some basic features of higher-order gravity, on the other side they are sufficiently simple to be easy to work with \cite{Sotiriou:2008rp}. Besides, there is one other important reason that convinces us that the $f(R)$ theories of gravity are good candidates to help us evaluate the inflationary observables based on the Tsallis entropy. These models can avoid serious problems such as negative energies and related instabilities that are called the Ostrogradski instabilities \cite{woodard2007avoiding, Sotiriou:2008rp, Ostrogradsky:1850fid}. This feature made the $f(R)$ theories unique compared to other higher-order gravity theories.

The paper is structured as follows. In Sec. \ref{section:thermodynamics:Tsallis}, we briefly study the thermodynamics of the Tsallis cosmology and the background equations derived from the first law of thermodynamics. In Sec. \ref{section:inflation:fR}, we review the inflationary dynamics and spectra of primordial perturbations in the $f(R)$ gravity. The thermodynamic behavior of the field equations in the $f(R)$ gravity will be discussed in Sec. \ref{section:thermodynamics:fR}. Then, in Sec. \ref{section:inflation:Tsallis}, we reconstruct an $f(R)$ model which is thermodynamically equivalent to the Tsallis model. Using this equivalence, we derive the equations of the inflationary observables in our Tsallis inflation scenario. In Sec. \ref{section:observational}, we examine the observational compatibility of the quadratic and natural potentials with the Planck 2018 data in the framework of the Tsallis inflation. Finally, we summarize our concluding remarks in Sec. \ref{section:conclusions}.


\section{Thermodynamics of the Tsallis cosmology}
\label{section:thermodynamics:Tsallis}

In \cite{Sheykhi:2018dpn}, it has been shown that by starting from the first law of equilibrium thermodynamics, $d E=T_h dS_h+W d\mathcal{V}$, at the apparent horizon of the FRW Universe and taking the entropy associated with the apparent horizon in the form of the Tsallis entropy \cite{tsallis2013black}, one can derive the modified Friedmann equations. Here, $E$ is the total energy content of the Universe, $T_h$ is the temperature of the apparent horizon, $W$ is the work density, and $\mathcal{V}$ is the volume inside the apparent horizon. For these quantities, we have \cite{Akbar:2006kj, hayward1999dynamic, Sheykhi:2018dpn}
\begin{align}
\label{Energy}
E& =\rho \, \mathcal{V} \, ,
\\
\label{Th}
T_h &=- \frac{1}{2\pi \tilde{r}_{A}}\left(1-\frac{\dot{\tilde{r}}_A}{2 H \tilde{r}_A}\right) \, ,
\\
\label{workdensity}
W &= \frac{1}{2}(\rho-p) \, ,
\\
\label{volume}
\mathcal{V}&=\frac{4}{3}\pi \tilde{r}_A^3 \, ,
\end{align}
where $\rho$ and $p$ represent respectively the energy density and pressure of all matter components in the Universe, and they  satisfy the following continuity equation
\begin{equation}
\label{continutyequation}
\dot{\rho}+3 H (\rho+p)=0 \, .
\end{equation}
In the above equations, $H\equiv \dot{a}/a $ denotes the Hubble parameter, $\tilde{r}_{A}$ indicates the radius of the apparent horizon, and the overdot indicates the derivative with respect to the cosmic time $t$. The horizon entropy is denoted by $S_h$, and it will be taken in the form of the Tsallis entropy which is a non-extensive generalization of Boltzmann–Gibbs entropy \cite{tsallis1988possible, tsallis2013black, tsallis2019black}
\begin{equation}
\label{Tsallisentropy}
S_h=\gamma A^{\beta} \, .
\end{equation}
Here, $A=4\pi \tilde{r}_A^2$ is the area of the apparent horizon and $\beta$ is a real parameter known as the Tsallis parameter that measures the degree of non-extensivity \cite{tsallis2013black}. In addition, $\gamma$ is an unknown constant, and several definitions have been presented for it in some literature for convenience \cite{Sheykhi:2018dpn, Lymperis:2018iuz, Nojiri:2019skr, Nojiri:2019itp, Asghari:2021lzu}. In our investigation, however, this parameter will be treated as an unknown parameter. It is clear that for $\beta=1$ and $\gamma=1/(4G)$, where $G$ is the Newton gravitational constant, the Tsallis entropy \eqref{Tsallisentropy} reduces to the Bekenstein-Hawking entropy \cite{bekenstein1973black}.

On the flat FRW background, the modified Friedmann equations based on the Tsallis entropy \eqref{Tsallisentropy} can be derived as follows \cite{Sheykhi:2018dpn}
\begin{align}
\label{FR1:Tsallis}
&H^{2(2-\beta)}=\frac{8\pi\,G_{{\rm eff}}}{3}\rho \, , \\
\label{FR2:Tsallis}
& (4-2\beta)\,\frac{\ddot{a}}{a}\,H^{2(1-\beta)}+(2\beta-1)H^{2(2-\beta)}=-8\pi\,G_{{\rm eff}}\,p \, ,
\end{align}
where the effective gravitational constant is defined as
\begin{equation}
\label{Geff}
G_{\rm eff}\equiv \left(\frac{1}{\gamma}\right)\left(\frac{2-\beta}{4 \beta}\right) (4\pi)^{1-\beta} \, .
\end{equation}
In this paper, we use the unit system in which $c=\hbar=\kappa_{B}=1$ where $c$ is the speed of light, $\hbar$ is the reduced Planck constant, and $\kappa_{B}$ is the Boltzmann constant. We define $\kappa^{2}=8\pi G=8\pi/m_P^{2}=1/{M_P}^{2}$ where $m_P$ is the Planck mass with a reduced value $M_P=m_P/\sqrt{8\pi}= (8\pi G)^{-1/2}$ which throughout this paper we take it equal to unity, $M_P =1$. From Eqs. \eqref{FR1:Tsallis}, \eqref{FR2:Tsallis}, and \eqref{Geff}, one can easily show that the standard Friedmann equations based on the Einstein gravity in which the entropy associated with the horizon follows the form of Bekenstein-Hawking entropy \cite{bekenstein1973black}, are recovered for $\beta=1$ and $\gamma=1/(4G)$, as expected. It is worth mentioning that although the author of \cite{Sheykhi:2018dpn} has defined $\gamma$ as $\gamma \equiv (2-\beta)(4\pi)^{1-\beta}/(4\beta \, G$), we do not follow this convention in the present work. For a spatially flat Universe, we have $\tilde{r}_A=1/H$, and so one can rewrite the area of the apparent horizon $A$ in terms of the Hubble parameter $H$ as
\begin{equation}
\label{AH}
A=\frac{4\pi}{H^2} \, .
\end{equation}
We assume that the matter-energy content of the Universe to be a scalar field $\phi$ in the form of a perfect fluid with the energy-momentum tensor $T^{\mu}{}^{(\phi)}_{\nu}= \mathrm{diag}\, (-\rho_{\phi}, p_{\phi}, p_{\phi}, p_{\phi})$. Here, $\rho_{\phi}$ and $p_{\phi}$ denote the energy density and pressure of the scalar field, respectively, and they are given by
\begin{align}
\label{rhophi}
& \rho_{\phi}=\frac{1}{2}\dot{\phi}^2+V(\phi) \, , \\
\label{pphi}
& p_{\phi}=\frac{1}{2}\dot{\phi}^2-V(\phi) \, .
\end{align}
In these equations, $V(\phi)$ is the potential energy of the scalar field. The energy density $\rho_{\phi}$ and pressure $p_{\phi}$ of the scalar field fulfill the continuity equation
\begin{equation}
\label{continutyequation-phi}
\dot{\rho}_{\phi}+3 H (\rho_{\phi}+p_{\phi})=0 \, .
\end{equation}
From this equation together with Eqs. \eqref{rhophi} and \eqref{pphi}, the equation of motion of the scalar field equation will be obtained as
\begin{equation}
\label{eq:FieldTsallis}
\ddot{\phi}+3 H \dot{\phi}+ V_{, \phi}=0 \, ,
\end{equation}
where $(, \phi)$ indicates the partial derivative with respect to the scalar field $\phi$. As we see, Eq. \eqref{eq:FieldTsallis} is the same as the one that is valid in the standard scenario based on the Einstein gravity.

To study the inflationary epoch in the Tsallis entropy-based scenario, we need to derive the fundamental relations governing the theory of cosmological perturbations. For this purpose, we need the action of the model. But we don't know the action of the model, and to overcome this problem, we try to reconstruct an $f(R)$ gravity model that is equivalent to our model from the thermodynamic point of view. In the $f(R)$ gravity, the horizon entropy is given by \cite{Akbar:2006mq, DeFelice:2010aj}
\begin{equation}\label{fRentropy}
S=\frac{A F(R)}{4G} \, ,
\end{equation}
where $F\equiv df(R)/dR$. In the following, we review the basic formulas governing the theory of cosmological perturbations in the $f(R)$ gravity, as well as the thermodynamic behavior of field equations in this theory. Then we apply the obtained results in our Tsallis entropy-based model and find the necessary relations to study inflation.


\section{Inflationary dynamics and spectra of primordial perturbations in the $f(R)$ gravity}
\label{section:inflation:fR}

The $f(R)$ gravity is described by the following action \cite{DeFelice:2010aj}
\begin{equation}
\label{action}
S= \int {\rm d}^{4}x \sqrt{-g}\left[\frac{1}{2\kappa^2} f(R) + X-V(\phi)\right] \, ,
\end{equation}
where $g$ is the determinant of the metric $ g_{{\mu}{\nu}}$, $f(R)$ is an arbitrary function of the Ricci scalar $R$, and $X\equiv -\frac{1}{2}g^{\mu\nu}\partial_\mu{\phi}\partial_\nu{\phi}$ is the canonical kinetic term. For the flat FRW metric, the Ricci scalar $R$ is given by \cite{DeFelice:2010aj}
\begin{equation}\label{R}
R=6\left(2H^2+\dot{H}\right) \, .
\end{equation}
For the $f(R)$ gravity model with action \eqref{action}, the Friedmann equations turn into \cite{DeFelice:2010aj}
\begin{align}
3FH^{2} & =\frac{1}{2}\left(RF-f\right)-3H\dot{F}+\kappa^{2}\left[\frac{1}{2}\dot{\phi}^{2}+V(\phi)\right] \, ,
\label{FR1-fR}
\\
-2F\dot{H} &= \ddot{F}-H\dot{F}+\kappa^{2}\dot{\phi}^{2} \, .
\label{FR2-fR}
\end{align}

Following \cite{Hwang:2001pu, DeFelice:2010aj}, we introduce the slow-roll parameters as
\begin{equation}
\label{SRparameters}
\varepsilon_{1} \equiv -\frac{\dot H}{H^2} \, , \hspace{.5cm} \varepsilon_{2}\equiv \frac{\ddot{\phi}}{ H\, \dot{\phi}} \, , \hspace{.5cm} \varepsilon_{3}\equiv \frac{\dot{F}}{2 H F} \, , \hspace{.5cm} \varepsilon_{4}\equiv \frac{\dot{E}}{2 H E} \, ,
\end{equation}
where
\begin{equation}\label{E}
E\equiv F\left(1+\frac{3 \dot{F}^2}{2 \kappa^2 \dot{\phi}^2 F}\right) \, .
\end{equation}
In the slow-roll regime, the quantities $\mid\varepsilon_{i}\mid \hspace{.1cm} (i=1, 2, 3, 4)$ are much smaller than unity.
Under the slow-roll limit, the Ricci scalar $R$ in Eq. \eqref{R} reduces to
\begin{equation}
\label{R:SR}
R\simeq 12 H^2.
\end{equation}

The spectrum of the curvature perturbations generated during inflation and in the slow-roll limit can be estimated as \cite{DeFelice:2010aj}
\begin{equation}
\label{Ps1}
{\cal P}_{s}\simeq\frac{1}{Q_s}\left(\frac{H}{2\pi}\right)^2 \, ,
\end{equation}
where
\begin{equation}
\label{Qs}
Q_s \simeq \dot{\phi}^2 \left(\frac{E}{F H^2}\right) \, .
\end{equation}
It is worth mentioning that ${\cal P}_{s}$ is computed at the time of horizon exit at which $k = aH$, where $k$ is the comoving wavenumber. The observational value of the amplitude of scalar perturbations at the CMB pivot scale $k_{*}=0.05\, {\rm Mpc}^{{\rm -1}}$ has been constrained by the Planck 2018 CMB observations to be ${\cal P}_{s}(k_{*})\simeq 2.1 \times 10^{-9}$ \cite{Planck:2018jri}.

The scalar spectral index $n_s$ during the slow-roll regime, in the framework of the $f(R)$ gravity is given by \cite{DeFelice:2010aj}
\begin{equation}
\label{nsSR}
n_s-1\simeq -4\varepsilon_{1}-2\varepsilon_{2}+2\varepsilon_{3}-2\varepsilon_{4} \, .
\end{equation}
The observational constraint from the Planck data on the scalar spectral index at the CMB pivot scale $k_{*}=0.05\, {\rm Mpc}^{{\rm -1}}$ is $n_{s}=0.9657\pm 0.0044$ (68\% CL, Planck 2018 TTTEEE+low$\ell$+lowE) \cite{Planck:2018jri}.

For $f(R)=R$, from Eqs. \eqref{SRparameters}, \eqref{E}, and \eqref{Qs}, it follows that $\varepsilon_{3}=0$, $\varepsilon_{4}=0$, and $Q_s=\dot{\phi}^2/H^2$. Substitution of these results into Eqs. \eqref{Ps1} and \eqref{nsSR} leads to ${\cal P}_{s}\simeq H^4/(4 \pi^2 \dot{\phi}^2)$ and $n_s-1\simeq -4\varepsilon_{1}-2\varepsilon_{2}$ which are the familiar relations in standard inflation model based on the Einstein gravity \cite{Liddle:1992wi, Stewart:1993bc}.

The tensor power spectrum in the framework of the $f(R)$ gravity is given by \cite{DeFelice:2010aj}
\begin{equation}
\label{Pt}
{\cal P}_{t}\simeq \frac{16}{\pi}\left(\frac{H}{m_P}\right)^2 \frac{1}{F} \, .
\end{equation}
Also, the tensor spectral index $n_{t}$ in this setup is obtained as
\begin{equation}
\label{ntSR}
n_t \simeq -2\varepsilon_{1}-2\varepsilon_{3} \, .
\end{equation}
This parameter determines the scale dependence of the tensor power spectrum. Currently, there is no precise measurement for this quantity and we hope the future observations can provide some constraints on this observable.

Using Eqs. \eqref{Ps1} and \eqref{Pt}, one can find the tensor-to-scalar ratio in the $f(R)$ gravity setting and in the slow-roll regime as \cite{DeFelice:2010aj}
\begin{equation}
\label{r}
r\equiv \frac{{\cal P}_{t}}{{\cal P}_{s}}\simeq \frac{64\pi}{m_P^{2}} \left(\frac{Q_s}{F}\right) \, .
\end{equation}
The Planck 2018 data sets an upper bound on the tensor-to-scalar ratio as $r<0.0522$ at the CMB pivot scale $k_{*}=0.05\, {\rm Mpc}^{{\rm -1}}$ (68\% CL, Planck 2018 TTTEEE+low$\ell$+lowE) \cite{Planck:2018jri}. The most recent upper limit on this parameter is $r_{0.01}<0.028$ at 95\% CL which is obtained at the CMB pivot scale $k=0.01\, {\rm Mpc}^{{\rm -1}}$ using 10 datasets from the BICEP/Keck Array 2015 and 2018, Planck releases 3 and 4, and LIGO-Virgo-KAGRA Collaboration \cite{Galloni:2022mok}.

In the case of $f(R)=R$, using Eqs. \eqref{Qs}, \eqref{Pt}, \eqref{ntSR}, and \eqref{r}, we easily find that ${\cal P}_{t}\simeq 2 H^2/\pi ^2$, $n_t \simeq -2\varepsilon_{1}$ and $r\simeq 16\varepsilon_{1}$. As we see, in this case, these results are reduced to the well-known relations in the standard inflation.


\section{Thermodynamic behavior of field equations in the $f(R)$ gravity}
\label{section:thermodynamics:fR}

Let us discuss the relation between the first law of thermodynamics and field equations in the $f(R)$ gravity. In \cite{Akbar:2006mq, DeFelice:2010aj}, it has been shown that the first law of equilibrium thermodynamics, $d E=T_h dS_h+W d\mathcal{V}$, does not hold at the apparent horizon of the FRW Universe in the $f(R)$ gravity. Consequently, to derive the Friedmann equations, we should apply the non-equilibrium thermodynamics and write the first law of thermodynamics as $d E=T_h dS_h+W d{\mathcal{V}}+ T_h d\bar{S}$. In this equation, $S_h$ is the entropy associated with the apparent horizon and it is still given by Eq. \eqref{fRentropy}. Furthermore, $\bar{S}$ implies the non-equilibrium entropy and involves the non-equilibrium thermodynamic effects of the $f(R)$ gravity \cite{Akbar:2006mq, DeFelice:2010aj}.

From the non-equilibrium relation $d E=T_h dS_h+W d{\mathcal{V}}+ T_h d\bar{S}$, we find
\begin{equation}
\label{dsbartods1}
\frac{ d\bar{S}}{dS_h}=-1+\frac{d E-W d{\mathcal{V}}}{T_h dS_h} \, .
\end{equation}
Taking the differentiation of Eq. \eqref{Energy}, and then using Eq. \eqref{volume} and also the relation $d{\mathcal{V}}=4\pi \tilde{r}_A^{2}d\tilde{r}_A$, we obtain
\begin{equation}
\label{dE1}
dE=4\pi \tilde{r}_A^2 \rho d\tilde{r}_A +\frac{4}{3}\pi \tilde{r}_A^{3} \dot{\rho}dt \, .
\end{equation}
With the help of Eq. \eqref{continutyequation}, we get
\begin{equation}
\label{dE2}
dE=4\pi \tilde{r}_A^2 \rho d\tilde{r}_A -4\pi H \tilde{r}_A^{3} (\rho+p)dt \, .
\end{equation}
It is assumed that the entropy of the horizon $S_h$ is in the form of Eq. \eqref{fRentropy}. Differentiating the entropy \eqref{fRentropy}, it follows that
\begin{equation}
\label{ds:fR}
dS_h=\frac{\pi}{G} \left(\tilde{r}_A^2 \dot{F} +2 F\tilde{r}_A \dot{\tilde{r}}_A\right)dt \, .
\end{equation}
where we have used $A=4\pi \tilde{r}_A^{2}$ in deriving the above equation.

Now, using Eqs. \eqref{Th}, \eqref{workdensity}, \eqref{volume}, \eqref{dE2}, and \eqref{ds:fR}, in the right side of Eq. \eqref{dsbartods1}, we find
\begin{equation}
\label{dsbartods2}
\frac{d\bar{S}}{dS_h}=\frac{8\pi G H \tilde{r}_A^3(\rho+p)-\tilde{r}_A \dot{F}-2 F \dot{\tilde{r}}_A}
{\tilde{r}_A \dot{F}+2F\dot{\tilde{r}}_A} \, .
\end{equation}
Supposing the matter-energy content of the Universe to be a scalar field $\phi$ with the energy density $\rho=\rho_{\phi}$ and the pressure $p=p_{\phi}$, we use Eqs. \eqref{rhophi} and \eqref{pphi}, and then we can rewrite Eq. \eqref{dsbartods2} as
\begin{equation}
\label{dsbartods3}
\frac{d\bar{S}}{dS_h}=\frac{- H \dot{F}+2F \dot{H}+8\pi G \dot{\phi}^2}{H \dot{F}-2 F \dot{H}} \, .
\end{equation}
Note that in deriving Eq. \eqref{dsbartods3}, we have used the relation $\tilde{r}_A=1/H$. Finally, with the help of the second and third relations in Eq. \eqref{SRparameters}, we find
\begin{equation}
\label{dsbartods4}
\frac{d\bar{S}}{dS_h}=-1+\frac{4\pi G \, \dot{\phi}^2}{F H^2 (\varepsilon_1+\varepsilon_3)} \, .
\end{equation}
From Eq. \eqref{dsbartods1}, it is clear that in the absence of the term $d\bar{S}/dS_h$ the first-law of equilibrium thermodynamics $d E=T_h dS_h+W d\mathcal{V}$ on the apparent horizon holds. Therefore, Eq. \eqref{dsbartods4} is an important relation in our examination. In the following, we show that in the framework based on the Tsallis entropy and under the slow-roll approximation, the term $d\bar{S}/dS_h$ vanishes, and hence $d\bar{S}=0$. Therefore, this point allows us to use the obtained relations in the $f(R)$ gravity to examine the inflationary models in the Tsallis entropy-based setting.


\section{Inflationary dynamics in the Tsallis entropy-based model}
\label{section:inflation:Tsallis}

In this section, we assume that the Tsallis entropy-based model and the $f(R)$ gravity are equivalent thermodynamically in the slow-roll regime.
With the help of Eq. \eqref{R:SR}, one can easily express the horizon area $A$ in Eq. \eqref{AH} in terms of Ricci scalar $R$ as
\begin{equation}
\label{AR}
A\simeq \frac{48\pi}{R} \, .
\end{equation}
Since we have supposed that the Tsallis entropy-based model and the $f(R)$ gravity are equivalent in the slow-roll limit, we set Eqs. \eqref{Tsallisentropy} and \eqref{fRentropy} equal to each other. Next, by using Eq. \eqref{AR}, we reach the following differential equation
\begin{equation}
\label{differentialEq}
\frac{12\pi}{G} F(R)-\gamma (48\pi)^{\beta}R^{1-\beta}\simeq 0 \, .
\end{equation}
Since $F(R)\equiv df(R)/dR$, solving this differential equation analytically, the function $f(R)$ is obtained as the following form
\begin{equation}
\label{fR}
f(R)=\left(\frac{4^{-1+2\beta}(3\pi)^{-1+\beta}\gamma \, G}{2-\beta}\right) R^{2-\beta} \, .
\end{equation}
As we see, $f(R)$ is a power-law function of $R$ as $f(R)=\mu R^n$ where $\mu\equiv (4^{-1+2\beta}(3\pi)^{-1+\beta}\gamma \, G)/(2-\beta)$ and $n\equiv2-\beta$. For $\beta=1$ and $\gamma=1/(4G)$, we find $f(R)=R$, and then Eq. \eqref{fRentropy} reduces to the Bekenstein-Hawking area-law of entropy \cite{bekenstein1973black} in the Einstein gravity.

Here, it is worthwhile to point out that the $R^n$ Lagrangian was originally regarded in the context of higher derivative theories \cite{Schmidt:1989zz, Maeda:1988ab}, and then applied to inflation \cite{Muller:1989rp, Gottlober:1992rg, Motohashi:2014tra}, which provides a simple and practical generalization of the Starobinsky $R^2$ inflation \cite{Starobinsky:1980te}. In particular, in \cite{Motohashi:2014tra}, this Lagrangian with $n \approx 2$ has been investigated to establish a way to measure a deviation from the $R^2$ inflation \cite{Starobinsky:1980te}. However, our methodology in the present paper differs from the approach of \cite{Motohashi:2014tra} in several aspects which are as follows. The action in our model is completely different from the action of \cite{Motohashi:2014tra} because, in our model, the gravitational part of the action consists of only the $R^n$ term, but the action of \cite{Motohashi:2014tra} contains the term $R + R^n$ which includes also the standard Einstein-Hilbert term. Furthermore, in our model, the contribution of a scalar field has been included in the action beside the $R^n$, but such a contribution is absent in the action of \cite{Motohashi:2014tra}. In the analysis of \cite{Motohashi:2014tra}, a conformal transformation from the Jordan frame to the Einstein frame has been performed, and the calculations of the inflationary observables are accomplished in the Einstein frame. In contrast, in our analysis, all calculations are performed in the Jordan frame. Finally, the form of the potentials that we consider in our work differs from the potentials that are regarded in \cite{Motohashi:2014tra} for the Einstein-frame scalar field.

Using Eq. \eqref{R:SR}, one can easily rewrite Eq. \eqref{fR} as the following form
\begin{equation}
\label{fRH}
f=\left(\frac{3 \, \pi^{-1+\beta} 4^{1+\beta} \gamma\,G}{2-\beta}\right)H^{2(2-\beta)} \, .
\end{equation}
From the slow-roll parameters \eqref{SRparameters}, we find that the field equations \eqref{FR1:Tsallis} and \eqref{eq:FieldTsallis} in the slow-roll limit lead to
\begin{align}
\label{FR1:Tsallis:SR}
&H^{2(2-\beta)}\simeq\frac{8\pi\,G_{{\rm eff}}}{3}V(\phi) \, ,
\\
\label{eq:FieldTsallis:SR}
& 3 H\dot{\phi}+V_{, \phi}\simeq 0 \, .
\end{align}
It can be shown that in the slow-roll approximation, the same equation as Eq. \eqref{FR1:Tsallis:SR} can also be derived by using the first Friedmann equation \eqref{FR1-fR} in the $f(R)$ gravity. This arises from the fact that in the slow-roll regime, the non-equilibrium entropy can be neglected in front of the equilibrium entropy, as we will show explicitly at the end of this section. Applying Eq. \eqref{FR1:Tsallis:SR}, we can rewrite the function $f$ in Eq. \eqref{fRH} in terms of the scalar field $\phi$ as
\begin{equation}
\label{fRphi}
f=\left(\frac{ 32\,G_{{\rm eff}} (4\pi)^{\beta} \gamma\,G}{2-\beta}\right)V(\phi) \, .
\end{equation}
Taking the time derivative of Eq. \eqref{FR1:Tsallis:SR} and then applying Eq. \eqref{eq:FieldTsallis:SR}, we can rewrite $\varepsilon_{1}$ in Eq. \eqref{SRparameters} as
\begin{equation}
\label{epsilon1}
\varepsilon_{1}\simeq \frac{1}{2(2-\beta)}\left(\frac{V_{, \phi}^2}{3 H^2 V(\phi)}\right) \, .
\end{equation}
For $\beta=1$ and $\gamma=1/(4G)$, from Eqs. \eqref{Geff}, \eqref{FR1:Tsallis:SR}, and \eqref{epsilon1}, one can easily verify that the corresponding relation in the standard slow-roll inflationary model will be recovered. It is known that inflation ends when the first slow-roll parameter $\varepsilon_{1}$ reaches unity \cite{martin3787encyclopaedia}.

Since $d/dt=\dot{\phi}d/{d\phi}$ and $d/dR=(1/{R_{, \phi}})d/d\phi$, we can also rewrite the slow-roll parameters $\varepsilon_{2}$, $\varepsilon_{3}$, and $\varepsilon_{4}$ in Eq. \eqref{SRparameters} in the following forms
\begin{align}
\label{epsilon2}
&\varepsilon_{2}\simeq \frac{H_{, \phi}V_{, \phi}-H V_{,\phi\phi}}{3 H^3} \, ,
\\
\label{epsilon3}
& \varepsilon_{3}\simeq \frac{V_{, \phi}\left(f_{, \phi}R_{, \phi\phi}-R_{, \phi}f_{, \phi\phi}\right)}{6 H^2 f_{, \phi}R_{, \phi}} \, ,
\\
\label{epsilon4}
& \varepsilon_{4}\simeq \frac{V_{, \phi}\left(f_{,\phi}R_{, \phi\phi}-R_{, \phi}f_{, \phi\phi}\right)\Big(R_{, \phi}^3+6 f_{, \phi} R_{, \phi\phi}^2+3R_{, \phi}^2 f_{, \phi\phi\phi}
-3R_{, \phi}(2f_{, \phi\phi}R_{, \phi\phi}+f_{, \phi}R_{, \phi\phi\phi})\Big)}{3 H^2 R_{, \phi}\Big(3 R_{, \phi}^2 f_{, \phi\phi}^2+3 f_{, \phi}^2 R_{, \phi\phi}^2+2 f_{, \phi}(R_{, \phi}^3-3 R_{, \phi} f_{, \phi\phi} R_{, \phi\phi})\Big)}.
\end{align}
In deriving Eqs. \eqref{epsilon2}, \eqref{epsilon3}, and \eqref{epsilon4}, we have also used Eqs. \eqref{E} and \eqref{eq:FieldTsallis:SR}. Besides, applying Eqs. \eqref{E}, \eqref{Qs}, and \eqref{eq:FieldTsallis:SR}, the scalar power spectrum ${\cal P}_{s}$ in Eq. \eqref{Ps1} takes the form
\begin{equation}
\label{Psphi}
{\cal P}_{s}\simeq \left(\frac{9 H^6}{2 \pi^2 V_{, \phi}^2}\right)\left(\frac{f_{, \phi}R_{, \phi}^3}{2 f_{, \phi}R_{, \phi}^3+3 (R_{, \phi}f_{, \phi\phi}-f_{, \phi}R_{, \phi\phi})^2}\right).
\end{equation}

Substituting Eqs. \eqref{epsilon1}, \eqref{epsilon2}, \eqref{epsilon3}, and \eqref{epsilon4} into Eq. \eqref{nsSR}, the scalar spectral index $n_s$ reads
\begin{align}\label{nsphi}
n_s \simeq & 1 +
\frac{1}{3 H^3}\Big[-2 H_{, \phi} V_{, \phi}+\frac{2 H V_{, \phi}^2}{(\beta-2)V} +H V_{, \phi}\left(\frac{R_{, \phi\phi}}{R_{, \phi}}-\frac{f_{, \phi\phi}}{f_{, \phi}}\right)+2 H V_{, \phi\phi}
\nonumber
\\
&
+
\Big(2 H V_{, \phi}(R_{, \phi}f_{, \phi\phi}-f_{, \phi}R_{, \phi\phi})(R_{, \phi}^3+6f_{, \phi}
R_{, \phi\phi}^2+3R_{, \phi}^2 f_{, \phi\phi\phi}
\nonumber
\\
&
-3 R_{, \phi}(2f_{, \phi\phi}R_{, \phi\phi}+ f_{, \phi}R_{, \phi\phi\phi})) \Big)\Big]
\Big/
\Big[R_{, \phi}(3 R_{, \phi}^2 f_{, \phi\phi}^2+ 3 f_{, \phi}^2 R_{, \phi\phi}^2
\nonumber
\\
&
+2 f_{, \phi}(R_{, \phi}^3-3 R_{, \phi}f_{, \phi\phi}R_{, \phi\phi}))\Big] \, .
\end{align}
We can also rewrite the tensor-to-scalar ratio $r$ in Eq. \eqref{r} in terms of the scalar field $\phi$ as
\begin{equation}
\label{rphi}
r \simeq \left(\frac{4 V_{, \phi}^2 }{9 H^4}\right)\left(\frac{R_{, \phi}\big(2 f_{, \phi} R_{, \phi}^3+3 (R_{, \phi}f_{, \phi\phi}-f_{, \phi}R_{, \phi\phi})^2\big)}{f_{, \phi}^2 R_{, \phi}^3}\right) \, ,
\end{equation}
where we have used Eqs. \eqref{E}, \eqref{Qs}, and \eqref{eq:FieldTsallis:SR} together with the relations $d/dt=\dot{\phi}d/{d\phi}$ and $d/dR=(1/{R_{, \phi}})d/d\phi$.

It is convenient to evaluate the inflationary observations in terms of the so-called $e$-fold number $N$ which measures the growth of the scale factor $a$ during inflation. It is defined as
\begin{equation}
\label{efolddefinition1}
N\equiv\ln\left(\frac{a_{\rm e}}{a}\right),
\end{equation}
where the subscript ``${\rm e}$'' refers to end of inflation. The definition \eqref{efolddefinition1} leads to
\begin{equation}
\label{efolddefinition2}
dN=-H dt=-\frac{H}{\dot{\phi}}d\phi \, .
\end{equation}
Note that the anisotropies observed in the CMB exit the Hubble horizon around $N_*\approx 50–60$ $e$-folds before the end of inflation \cite{Dodelson:2003vq, Liddle:2003as}. The precise value of horizon exit $e$-fold number $N_*$ depends on the energy scale of inflation and also on the details of the reheating process after inflation \cite{Dodelson:2003vq, Liddle:2003as}. In our model, like most conventional inflationary models, the features of the reheating mechanism after inflation are unknown to us, and therefore it is not possible to determine the precise value of $N_{*}$.

Using the last equality in Eq. \eqref{efolddefinition2} and also applying Eqs. \eqref{FR1:Tsallis:SR} and \eqref{eq:FieldTsallis:SR}, we reach the following differential equation
\begin{equation}
\label{dphitodN:SR}
\frac{d\phi}{dN}\simeq \left( 3^{1-\beta}8\pi G_{\rm eff}\right)^\frac{1}{\beta-2} \left(\frac{V_{, \phi}}{V^{1/(2-\beta)}}\right) \, .
\end{equation}
One can solve Eq. \eqref{dphitodN:SR} to obtain the scalar field $\phi$ as a function of the $e$-fold number $N$ in the slow-roll approximation. In this way, we find the inflationary observables in terms of $N$.

Using Eqs. \eqref{FR1:Tsallis:SR} and \eqref{eq:FieldTsallis:SR}, we can rewrite Eq. \eqref{dsbartods4} in the slow-roll limit as
\begin{equation}
\label{dsbartods1:SR}
\frac{d\bar{S}}{dS_h}\simeq-1+3^\frac{2}{2-\beta}\left( \frac{4\pi G \, }{9 F (8\pi G_{\rm eff})^\frac{2}{2-\beta}(\varepsilon_1+\varepsilon_3)}\right)\left(\frac{V_{, \phi}}{V^{1/(2-\beta)}}\right)^2 \, .
\end{equation}
Using Eq. \eqref{R:SR}, we obtain $R_{, \phi}=24 H H_{, \phi}$, and since $F=df/dR=f_{,\phi}/R_{,\phi}$, we have $F=f_{, \phi}/(24 H H_{, \phi})$. Substituting this relation into Eq. \eqref{dsbartods1:SR}, we get
\begin{equation}
\label{dsbartods2:SR}
\frac{d\bar{S}}{dS_h}\simeq -1+3^\frac{2}{2-\beta}\left( \frac{96\pi G \, H_{, \phi}H}{9 f_{, \phi}(8\pi G_{\rm eff})^\frac{2}{2-\beta}(\varepsilon_1+\varepsilon_3)}\right)\left(\frac{V_{, \phi}}{V^{1/(2-\beta)}}\right)^2 \, .
\end{equation}
Now substituting Eqs. \eqref{epsilon1} and \eqref{epsilon3} into Eq. \eqref{dsbartods2:SR}, and then using Eqs. \eqref{Geff}, \eqref{FR1:Tsallis:SR}, and \eqref{fRphi}, we will have
\begin{equation}
\label{dsbar}
\frac{d\bar{S}}{dS_h}\simeq 0 \hspace{0.5cm} \Rightarrow \hspace{0.5cm} d\bar{S}\simeq 0 \, ,
\end{equation}
which means that in the slow-roll Tsallis entropy-based inflation, the first law of equilibrium thermodynamics $d E=T_h dS_h+W d\mathcal{V}$ on the apparent horizon holds. This point makes us sure to apply the results derived in the $f(R)$ gravity to find the inflationary observables in the Tsallis entropy-based cosmology.


\section{Observational constraints}
\label{section:observational}

In this section, we apply the obtained results in the previous section to investigate the observational consistency of two different inflationary potentials in the framework of the Tsallis entropy-based inflation. These potentials are the quadratic and natural potentials that are not in good agreement with the current CMB data in the setting of standard inflation.


\subsection{Quadratic potential}
\label{subsection:quadratic}

Let us continue studying the inflationary scenario in the Tsallis entropy-based setting by considering the quadratic potential \cite{Linde:1983gd}
\begin{equation}
\label{potential:phi2}
V(\phi)=\frac{1}{2} m^{2}\phi^{2} \, ,
\end{equation}
where $m$ is the inflaton mass.
Applying the potential \eqref{potential:phi2} and also using Eqs. \eqref{Geff}, \eqref{R:SR}, \eqref{FR1:Tsallis:SR}, and \eqref{fRphi} , the slow-roll parameters in Eqs. \eqref{epsilon1}, \eqref{epsilon2}, \eqref{epsilon3}, and  \eqref{epsilon4} take the form
\begin{align}
\label{epsilon1:phi2}
&\varepsilon_{1}\simeq \left(\frac{3^{\frac{-1+\beta}{2-\beta}}}{2-\beta}\right) \left(\frac{4^{1-\beta}\pi^{2-\beta}(2-\beta)}{\beta} \phi^2 \right)^{\frac{1}{\beta-2}} \Theta \, ,
\\
\label{epsilon1-2:phi2}
& \varepsilon_{2}=\varepsilon_{3}\simeq (-1+\beta) \left(\frac{3^{\frac{-1+\beta}{2-\beta}}}{2-\beta}\right)\left(\frac{4^{1-\beta}\pi^{2-\beta}(2-\beta)}{\beta} \phi^2 \right)^{\frac{1}{\beta-2}} \Theta \, ,
\\
\label{epsilon4:phi2}
&\varepsilon_{4}\simeq \left(\frac{3^{\frac{-1+\beta}{2-\beta}}}{2-\beta}\right) (-1+\beta)
\left(\frac{4^{1-\beta}\pi^{2-\beta}(2-\beta)}{\beta} \phi^2 \right)^{\frac{1}{\beta-2}} \Theta
\Bigg\{(12 \pi)^\beta (-1+\beta)\beta \Theta +4\pi^2(-2+\beta)^2
\nonumber\\
&
\times 3^{-\frac{(1-\beta)^2}{2-\beta}}
\left(\frac{4^{1-\beta}\pi^{2-\beta}(2-\beta)}{\beta} \phi^2 \right)^{\frac{-1+\beta}{2-\beta}}
\phi^2
\Bigg\}
\Big/ \Bigg\{(12 \pi)^\beta (-1+\beta)^2 \Theta+4\pi^2(-2+\beta)^2\,
  3^{-\frac{(1-\beta)^2}{2-\beta}}
 \nonumber\\
 &
 \times \left(\frac{4^{1-\beta}\pi^{2-\beta}(2-\beta)}{\beta} \phi^2 \right)^{\frac{-1+\beta}{2-\beta}}\phi^2
\Bigg\}.
\end{align}

The parameter $\Theta$ in these equations is defined as
\begin{equation}
\label{Theta}
\Theta \equiv m^2 \left(\frac{m^2}{\gamma} \right)^\frac{1}{\beta-2} \, .
\end{equation}
It is easy to show that for $\beta=1$, we have $\Theta=\gamma$. Therefore, for $\beta=1$ and $\Theta=1/(4G)$, we find the same relations in the Einstein gravity.

Using Eqs. \eqref{Geff}, \eqref{R:SR}, \eqref{FR1:Tsallis:SR}, \eqref{fRphi}, and \eqref{potential:phi2}, the scalar power spectrum \eqref{Psphi} turns into
\begin{align}
\label{Ps:phi2}
{\cal P}_{s}\simeq & \frac{3^{\frac{1-2\beta}{2-\beta}}m^2(2-\beta)^2 \Theta^{-3}\left(\frac{4^{1-\beta}\pi^{2-\beta}(2-\beta)}{\beta} \, \phi^2 \right)^{\frac{3}{2-\beta}}}{3^\frac{1}{2-\beta}(4\pi)^\beta (1-\beta)^2 \Theta \left(\frac{4^{1-\beta}\pi^{2-\beta}(2-\beta)}{\beta} \, \phi^2 \right)^{\frac{1-\beta}{2-\beta}}+4 \pi^2(2-\beta)^2 \phi^2} \, .
\end{align}
With the help of Eqs. \eqref{Geff}, \eqref{R:SR}, \eqref{FR1:Tsallis:SR}, \eqref{fRphi}, and \eqref{potential:phi2}, one can rewrite the scalar spectral index in Eq. \eqref{nsphi} and the tensor-to scalar ratio in Eq. \eqref{rphi} as
\begin{align}
n_s \simeq & 1 + \,  2 \, \Theta^{-\beta+2}\, 3^{\frac{1-\beta}{-2+\beta}}\Big[ 2^{1+2\beta} 3^{1+\beta}\pi^{\beta} +
3^{1+\beta}(4\pi)^{\beta} (-3+\beta^2) \beta+ (4\pi^2) (2-\beta)^{2} \, 3^\frac{-1-\beta+\beta^2}{-2+\beta}
\nonumber
\\
&
\times (1+\beta)\, \Theta^{-1}\left(\frac{4^{1-\beta}\pi^{2-\beta}(2-\beta)}{\beta} \, \phi^2 \right)^\frac{\beta-1}{2-\beta}
\phi^2
\Big]
\Big/
(\beta-2)\Big[3^{1+\beta}(4\pi)^\beta (1-\beta)^2
\nonumber
\\
&
\times \Theta^{-\beta+1}\left(\frac{4^{1-\beta}\pi^{2-\beta}(2-\beta)}{\beta} \, \phi^2 \right)^\frac{1}{2-\beta}+
(4\pi^2)\, 3^{\frac{1-\beta+\beta^2}{-2+\beta}} (2-\beta)^2
\Theta ^{-\beta}
\nonumber
\\
&
\times
\left(\frac{4^{1-\beta}\pi^{2-\beta}(2-\beta)}{\beta} \, \phi^2 \right)^\frac{\beta}{2-\beta} \phi^2
\Big] \, ,
\label{ns:phi2}
\end{align}
\begin{align}
r\simeq & \, 4\beta\, (2-\beta)^{-3} \, 3^{\frac{2-\beta^2}{\beta-2}} \, \Theta \left(\frac{4^{1-\beta}\pi^{2-\beta}(2-\beta)}{\beta} \, \phi^2 \right)^{-\frac{\beta}{2-\beta}}
\nonumber
\\
& \times
\left(3^{1+\beta}(4\pi)^{\beta}\pi^{-2}(1-\beta)^2 \Theta \, \phi^{-2}+4(2-\beta)^2 \, 3^\frac{\beta(1-\beta)+1}{2-\beta}\left(\frac{4^{1-\beta}\pi^{2-\beta}(2-\beta)}{\beta} \, \phi^2 \right)^{\frac{\beta-1}{2-\beta}} \right) \, .
\label{r:phi2}
\end{align}
Using Eqs. \eqref{Geff} and  \eqref{potential:phi2} in Eq. \eqref{fRphi}, we find
\begin{equation}\label{fphi:phi2}
f=\frac{2m^2}{\beta}\phi^2 \, .
\end{equation}
We can also find the Hubble parameter $H$ by substituting Eqs.  \eqref{Geff} and \eqref{potential:phi2} in Eq. \eqref{FR1:Tsallis:SR}, as
the following form
\begin{equation}\label{Hphi:phi2}
H=3^{\frac{1}{2(-2+\beta)}} m \Theta^{-\frac{1}{2}}
\left(\frac{2^{2(1-\beta)}\pi^{2-\beta}(2-\beta)\phi^2}{\beta}\right)^{\frac{1}{2(2-\beta)}} \, .
\end{equation}

With the help of Eq. \eqref{epsilon1:phi2} and solving the equation $\varepsilon_{1}(\phi_{\rm e})=1$, we can obtain the value of the scalar field at the end of inflation as
\begin{equation}
\label{phiend:phi2}
\phi_{\rm e}^{2}\simeq\frac{12^{-1+\beta}\,\pi^{-2+\beta}\beta\,\Theta^{2-\beta}}{(2-\beta)^{3-\beta}} \, .
\end{equation}
If we use Eqs. \eqref{Geff} and \eqref{potential:phi2} in Eq. \eqref{dphitodN:SR}, we will have
\begin{equation}
\label{dphitodN:phi2}
\frac{d\phi}{dN}\simeq 3^\frac{-1+\beta}{2-\beta}\, \Theta \left(\frac{4^{1-\beta}\pi^{2-\beta}(2-\beta)}{\beta} \, \phi^2 \right)^\frac{1}{\beta-2}\phi.
\end{equation}
Solving the differential equation \eqref{dphitodN:phi2}, we find the scalar field in terms of the $e$-fold number $N$ as
\begin{equation}
\label{phiN:phi2-1}
\phi(N)\simeq (2\pi)^{\frac{\beta}{2}} \left(\frac{\beta}{\pi^2(2-\beta)}\right)^{\frac{1}{2}}
\left(\frac{(2-\beta)}{3^{\frac{1-\beta}{\beta-2}}N\,\Theta+\left(\frac{(-1)^{1+\beta}\,2^{\beta}\pi^{-2+\beta}\,(\beta-2)^{\beta-3}\beta}
{\phi_{\rm e}^{2}}\right)^{\frac{1}{\beta-2}}}\right)^{\frac{\beta-2}{2}}.
\end{equation}
Substituting $\phi_e$ from Eq. \eqref{phiend:phi2} into Eq. \eqref{phiN:phi2-1}, we find
\begin{equation}
\label{phiN:phi2}
\phi(N)\simeq (2\pi)^{\frac{\beta}{2}}\left(\frac{\beta}{\pi^2(2-\beta)}\right)^{\frac{1}{2}}\left(\frac{3^{\frac{-1+\beta}{-2+\beta}}(2-\beta)}{\frac{1}{2}(2N+1)\Theta}\right)^{\frac{\beta-2}{2}}.
\end{equation}
When $\beta=1$ and $\Theta=1/(4G)$, Eq. \eqref{phiN:phi2} reduces to the obtained result in conventional inflation, i.e. $\phi(N)=\sqrt{2(2 N + 1)}$.

With the help of Eq. \eqref{phiN:phi2}, we can rewrite Eq. \eqref{fphi:phi2} in terms of the $e$-fold number $N$ and the parameters $\beta$, $\Theta$, and $m$ as
\begin{equation}\label{fN:phi2}
f=\frac{(2-\beta)^{-3+\beta}}{6\pi^2}(12\pi)^{\beta} m^2 \Big(\Theta (1+2N)\Big)^{2-\beta} \, .
\end{equation}
Applying Eq. \eqref{phiN:phi2} in Eq. \eqref{Hphi:phi2}, we take
\begin{equation}\label{HN:phi2}
H=\frac{m\sqrt{1+2N}}{\sqrt{3(2-\beta)}} \, .
\end{equation}
Now, we can substitute Eq. \eqref{phiN:phi2} into Eqs. \eqref{epsilon1:phi2}, \eqref{epsilon1-2:phi2}, and \eqref{epsilon4:phi2}, and obtain the slow-roll parameters as the following forms
\begin{align}
\label{epsilon1N:phi2}
&\varepsilon_{1}\simeq \frac{1}{1+2N},
\\
&
\label{epsilon2-3N:phi2}
\varepsilon_{2}=\varepsilon_{3}\simeq \frac{-1+\beta}{1+2N} \, ,
\\
&
\label{epsilon4N:phi2}
\varepsilon_{4}\simeq \frac{\beta(-1+\beta)\Big(3(-1+\beta)+(1+2N) \Big)}{(1+2N)\Big( 3(-1+\beta)^2+ \beta(1+2N)\Big)}.
\end{align}
Using Eq. \eqref{phiN:phi2} into Eq. \eqref{Ps:phi2}, we find the scalar power spectrum as
\begin{equation}
\label{PsN:phi2}
{\cal P}_s \simeq \frac{4\pi^{-\beta}(6-3\beta)^{-\beta} m^2 (1+2 N)^{2+\beta}}{3 (1-\beta)^2+\beta (1+2 N)}\,
\Theta^{\beta-2}.
\end{equation}
As we see from Eq. \eqref{PsN:phi2}, the power spectrum of the curvature perturbation is a function of the $e$-fold number $N$ and three free parameters $\beta$, $\Theta$, and $m$. One can use Eq. \eqref{PsN:phi2} and then impose the CMB normalization at the observable scale to find a constraint on the parameter $m$.

Applying Eq. \eqref{phiN:phi2}, the scalar spectral index \eqref{ns:phi2} and the tensor-to-scalar ratio \eqref{r:phi2} take the forms
\begin{align}
\label{nsN:phi2}
n_s & \simeq 1 -\frac{2\Big(6+3\beta (-3+\beta^2)+\beta (1+\beta)(1+2N)\Big)} {(1+2N)\Big(\beta (1+2N)+3(1-\beta)^2\Big)} \, ,
\\
\label{rN:phi2}
r & \simeq \frac{16 \Big(3(1-\beta)^2+\beta (1+2 N) \Big)}{(1+2 N)^2} \, .
\end{align}
In the limit of standard inflation where $\beta=1$, Eqs. \eqref{nsN:phi2} and \eqref{rN:phi2} reduces to $n_{s}=1-4/(2N+1)$ and $r=16/(2N+1)$, respectively. These results are the same equations that we find in the setup of the standard inflation. Let us now consider the limits where $\beta \ll 1$ and $N \gg 1$. From Eqs. \eqref{nsN:phi2} and \eqref{rN:phi2}, the leading contributions to $n_s$ and $r$ become
\begin{align}
\label{Starobinsky}
n_s\simeq & 1-\frac{2}{N} \, ,
\\
r\simeq & \frac{12}{N^2} \, .
\end{align}
This means that in the regime $\beta \ll 1$ and $N \gg 1$, our theory reduces to the $R^2$ inflationary model proposed by  Starobinsky \cite{Starobinsky:1980te}. In this regime, the power spectrum \eqref{PsN:phi2} reduces to the form
\begin{equation}
\label{PsN:phi2:c}
{\cal P}_{s}\simeq\frac{m^{2}(2N)^{2}}{3\Theta^{2}} \, .
\end{equation}
Using Eq. \eqref{PsN:phi2:c} and then impose the CMB normalization at the pivot scale $k_{*}=0.05\, {\rm Mpc}^{{\rm -1}}$ \cite{Planck:2018jri}, we find
\begin{align}
\label{m1}
m \simeq 7.937 \times 10^{-7} \Theta \hspace{1cm} {\rm for} \hspace{1cm} N_{*}=50 \, ,
 \\
m \simeq 6.614 \times 10^{-7} \Theta  \hspace{1cm} {\rm for} \hspace{1cm} N_{*}=60 \, .
\end{align}
As we see the value of the parameter $m$ depends on what value the parameter $\Theta$ takes.

\begin{figure}[t]
\begin{center}
\scalebox{0.9}[0.9]{\includegraphics{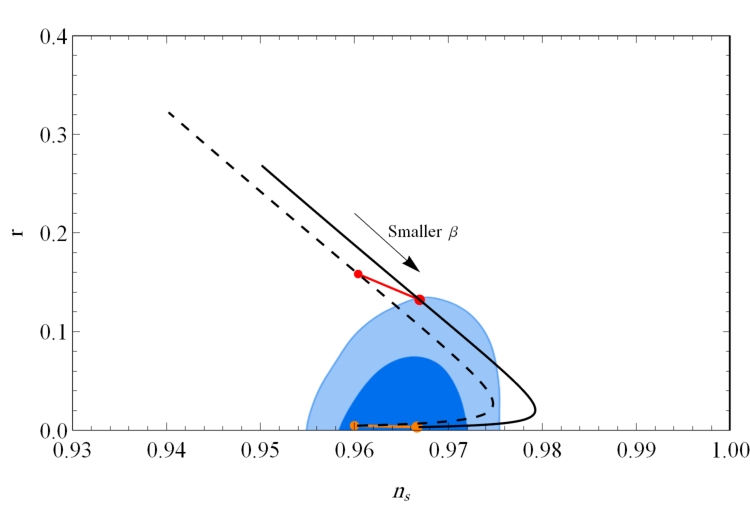}}
\caption{The $r-n_s$ diagram for the quadratic potential \eqref{potential:phi2} in the slow-roll inflation based on the Tsallis entropy for two different values of $N_*$ with varying $\beta$ in the range of $0<\beta<2$. The results for $N_*=50$ and $N_*=60$ are shown by the dashed and solid black curves, respectively. The red solid line between the dashed and solid black curves shows the results of the potential in the standard inflation, in the range of $50\leq N_* \leq 60$. Moreover, the result of the Starobinsky $R^2$ inflation \cite{Starobinsky:1980te} is specified by the orange solid line with $50\leq N_* \leq 60$. The marginalized joint 68\% and 95\% CL regions of the Planck 2018 TTTEEE+low$\ell$+lowE data \cite{Planck:2018jri} are specified by dark and light blue, respectively.}
\label{rns:phi2}
\end{center}
\end{figure}

With the help of Eqs. \eqref{nsN:phi2} and \eqref{rN:phi2}, we can plot the $r-n_{s}$ diagram and compare the prediction of our model with the Planck 2018 CMB data \cite{Planck:2018jri}. Fig. \ref{rns:phi2} shows the prediction of our model in the $r-n_{s}$ plane for two typical values of $N_*$ and varying $\beta$ in the range of $0<\beta<2$. The dashed and solid black curves illustrate the results of the model for $N_*=50$ and $N_*=60$, respectively. Besides, The red solid line between the dashed and solid black curves shows the prediction of the potential in the standard inflation which corresponds to $\beta=1$ in our model, in the range of $50\leq N_* \leq 60$. Moreover, the prediction of the Starobinsky $R^2$ inflationary model \cite{Starobinsky:1980te} has been shown by the orange solid line with $50\leq N_* \leq 60$.

In Fig. \ref{rns:phi2}, the parameter $\beta$ has been taken as a varying parameter in the range $0<\beta<2$. It should be noted that each value of $\beta$ in this range is related to a special case of our power-law $f(R)$ scenario, and in the case with $\beta = 1$ and $\gamma =1/(4G)$, our $f(R)$ model reduces to the Einstein general relativity (GR). This does not mean that a transition from $f(R)$ to GR has occurred during inflation in our scenario. In other words, to investigate each case of our scenario, we should fix the value of $\beta$ at the first step, and it is not the case that this parameter varies during inflation and causes a transition from $f(R)$ to GR. Since we cannot determine the parameter $\gamma$ in our investigation, therefore we cannot determine the time at which the $f(R)$ gravity in Eq. \eqref{fR} transits to GR.

Fig. \ref{rns:phi2} shows that the prediction of the potential $m^2 \phi^2/2$ in the standard setting is not in good consistency with the Planck 2018 observations \cite{Planck:2018jri}, while in the framework of the Tsallis entropy-based inflationary scenario, it can be in very good agreement with these data, and its results can lie inside the 68\% CL region of Planck 2018 data \cite{Planck:2018jri}. From the figure, we see that for small $\beta$, the model shows better consistency with the observations and the prediction of the model can enter the 68\% CL region of these data. As we have proved, in the limit $\beta\ll 1$, the inflationary observables $n_s$ and $r$ approach to the same values in the Starobinsky $R^2$ inflation \cite{Starobinsky:1980te}.

With the help of Eqs. \eqref{nsN:phi2} and \eqref{rN:phi2}, and also the Planck observational constraints on the $r–n_{s}$ plane, we can estimate the ranges of the parameter $\beta$ for which the results of the model in the $r-n_{s}$ plane are consistent with the 68\% CL region of the Planck 2018 data \cite{Planck:2018jri}. In the case $N_*=50$, if $0< \beta\lesssim 0.045$ the result of the model is in agreement with the 68\% CL constraint of these data, and for $N_*=60$ the prediction of our model can enter the 68\% CL region of the Planck 2018 data \cite{Planck:2018jri}, provided that $ 0<\beta\lesssim 0.011$.

\begin{figure}[t]
\begin{center}
\scalebox{0.9}[0.9]{\includegraphics{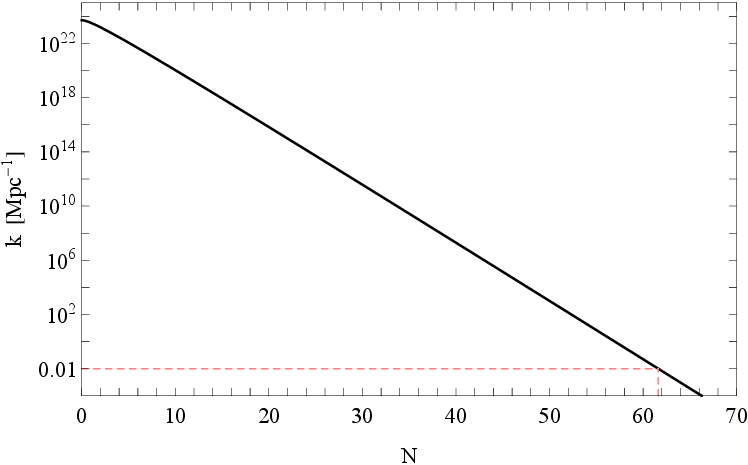}}
\caption{Evolution of the comoving wavenumber $k$ versus the $e$-fold number $N$ for the quadratic potential \eqref{potential:phi2} in the Tsallis inflationary setting. The pink horizontal and vertical dashed lines specify $k=0.01 \,  {\rm Mpc}^{{\rm -1}}$ and its corresponding $e$-fold number $N_{0.01} \simeq 61.62$, respectively.}
\label{k:phi2}
\end{center}
\end{figure}

Here, we are interested in applying the recent constraint of \cite{Galloni:2022mok} on $r_{0.01}$ to present some observational constraint on the model parameter $\beta$. For this purpose, we should determine the $e$-fold number at which the comoving wavenumber $k=0.01\,\mathrm{Mpc}^{-1}$ exits the Hubble horizon during inflation. To do so, we examine the behavior of the comoving wavenumber $k$ as a function of the $e$-fold number $N$, at which the mode with comoving wavenumber $k$ leaves the Hubble horizon, $k=a H$. With the help of this relation, we can easily find
\begin{equation}\label{kdefin}
k(N)=\left(\frac{a(N) H(N)}{a(N_*)H(N_*)}\right)k_{*},
\end{equation}
which $a(N_*)$ and $H(N_*)$ are the scale factor and the Hubble parameter at the time of horizon exit of the mode $k_{*}=0.05\,\mathrm{Mpc}^{-1}$, respectively. Using Eq. \eqref{efolddefinition1}, we can obtain the scale factor $a$ in terms of the $e$-fold number $N$ as
\begin{equation}\label{aN}
a(N)\simeq \exp({N_{*}-N}),
\end{equation}
where we have normalized the scale factor to its value at the epoch of horizon crossing of the mode $k_{*}=0.05\,\mathrm{Mpc}^{-1}$. Finally, applying Eqs. \eqref{HN:phi2} and \eqref{aN} in Eq. \eqref{kdefin}, the comoving number $k$ can be found as a function of  $N$ as
\begin{equation}\label{kN:phi2}
k(N)=\left(\sqrt{\frac{1+2N}{1+2N_{*}}}\exp({N_{*}-N})\right)0.05 \, {\rm Mpc}^{{\rm -1}} \, .
\end{equation}
Using Eq. \eqref{kN:phi2} and setting $N_* = 60$, we plot in Fig. \ref{k:phi2} the variation of the comoving wavenumber $k$ against the $e$-fold number $N$. In the figure, we have also specified the comoving wavenumber $k=0.01 \,  {\rm Mpc}^{{\rm -1}}$ and its corresponding $e$-fold number which is $N_{0.01} \simeq 61.62$.

\begin{figure}[t]
\begin{center}
\scalebox{0.9}[0.9]{\includegraphics{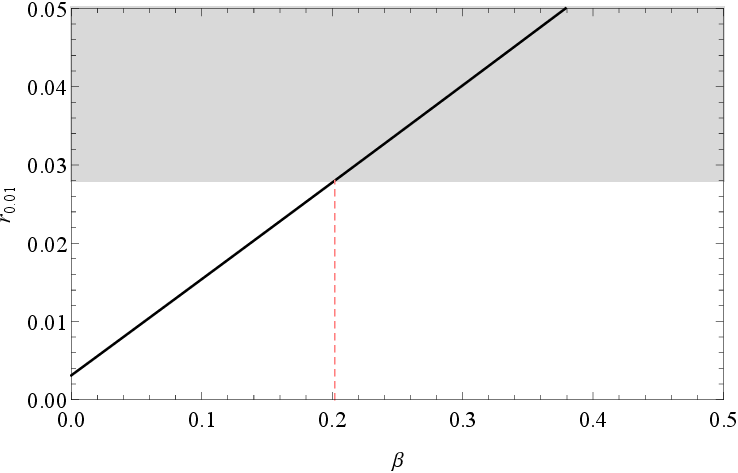}}
\caption{Variation of the tensor-to-scalar ratio $r_{0.01}$ versus the parameter $\beta$ by setting $N_{0.01} \simeq 61.62$, for the quadratic potential \eqref{potential:phi2} in the framework of the Tsallis inflation. The gray-shaded region is excluded by the constraint on the upper bound on $r_{0.01}$, reported by Galloni et al. \cite{Galloni:2022mok}. The pink vertical dashed line
specifies $\beta \simeq 0.202$ which is the maximum value of the parameter $\beta$ for which the model satisfies the constraint $r_{0.01}<0.028$.}
\label{rbeta:phi2}
\end{center}
\end{figure}

In Fig. \ref{rbeta:phi2}, with the help of Eq. \eqref{rN:phi2}, the variation of the tensor-to-scalar ratio $r_{0.01}$ is plotted as a function of the parameter $\beta$ by taking $N_{0.01} \simeq 61.62$. The gray shaded region is excluded by the constraint reported by Galloni et al. \cite{Galloni:2022mok} at the CMB pivot scale $k=0.01\, {\rm Mpc}^{{\rm -1}}$. Our results imply that the model satisfies the constraint $r_{0.01}<0.028$, provided that $0<\beta \lesssim 0.202$.

Using Eqs. \eqref{epsilon1N:phi2}, \eqref{epsilon2-3N:phi2}, and \eqref{epsilon4N:phi2}, we plot the evolution of the slow-roll parameters $\varepsilon_{1}$, $\varepsilon_{2}$, $\varepsilon_{3}$, and $\varepsilon_{4}$ as a function of $N$ for $\beta=0.01$ in Fig. \ref{SRparameters:phi2}. From the figure, it is obvious that the slow-roll approximation is valid until the end of inflation. This point proves the viability of our examination which is based on the slow-roll approximation. Besides, the figure shows that $\varepsilon_{1}$  becomes unity at $N = 0$ which corresponds to the end of inflation.

\begin{figure}[t]
\begin{center}
\scalebox{0.9}[0.9]{\includegraphics{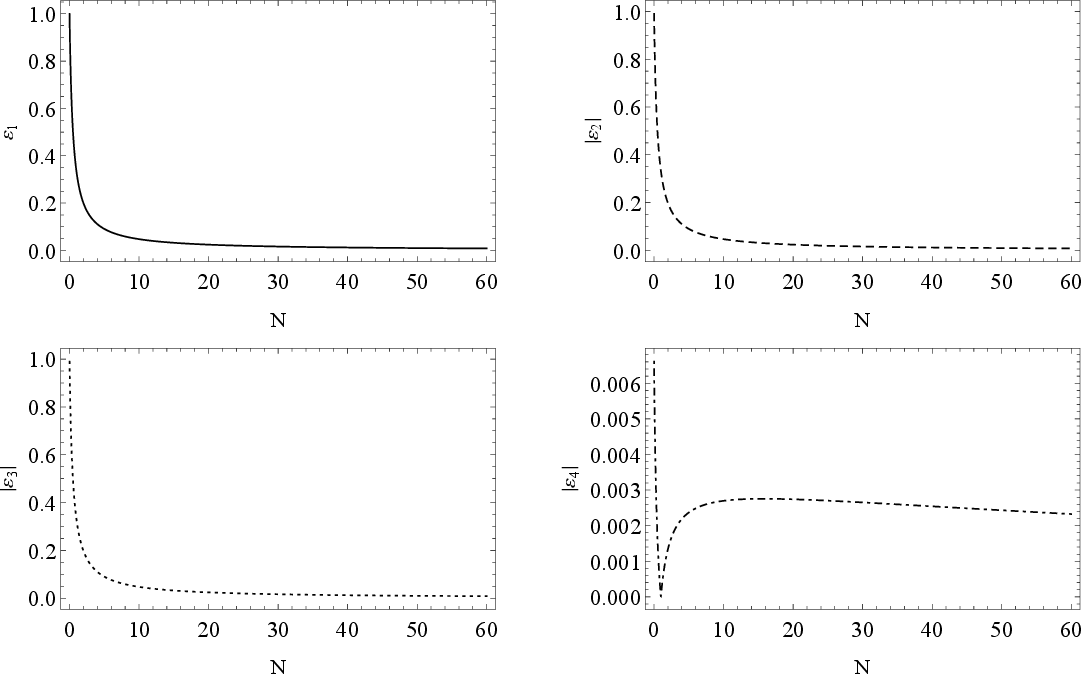}}
\caption{Evolution of the slow-roll parameters $\varepsilon_{1}$, $\varepsilon_{2}$, $\varepsilon_{3}$, and $\varepsilon_{4}$ versus the $e$-fold number $N$ for the quadratic potential \eqref{potential:phi2} in the setup of the Tsallis inflation, with $\beta=0.01$.}
\label{SRparameters:phi2}
\end{center}
\end{figure}

\begin{figure}[t]
\begin{center}
\scalebox{0.9}[0.9]{\includegraphics{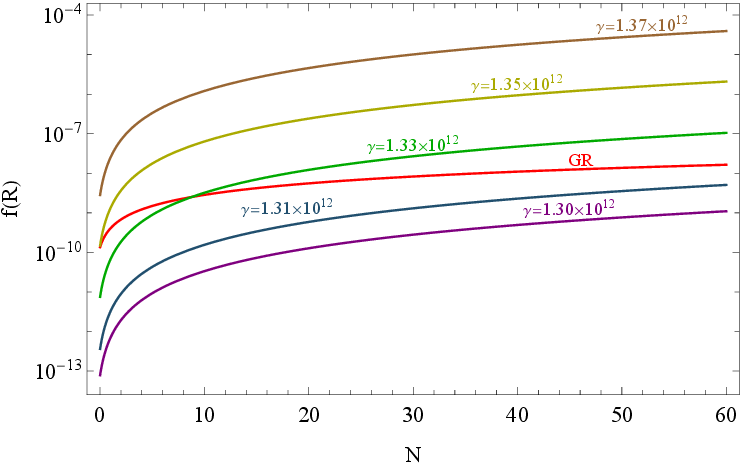}}
\caption{Evolution of the function $f(R)$ versus the $e$-fold number $N$ for the quadratic potential \eqref{potential:phi2} in the Tsallis inflationary scenario with $\beta=0.01$ and some typical values of $\gamma$. Also, the red curve illustrates the Ricci scalar $R$ predicted by general relativity (GR).}
\label{fR:phi2}
\end{center}
\end{figure}

Here, it should be noted that throughout this paper, we have worked in the framework of the $f(R)$ gravity given by Eq. \eqref{fR}, and we have treated GR as a special case of our $f(R)$ gravity model that is realized by taking $\beta = 1$ and $\gamma = 1/(4G)$. To clarify this point, we note that for a given set of model parameters $(\beta, \gamma)$, one can find $\Theta$ from Eq. \eqref{Theta} as a function of the parameter $m$. Substituting $\Theta=\Theta(m)$ in Eq. \eqref{PsN:phi2} and then fixing the amplitude of scalar perturbations ${\cal P}_s$ at the epoch of horizon crossing with the $e$-fold number $N_{*}=60$ as ${\cal P}_{s}(k_{*})\simeq 2.1 \times 10^{-9}$ \cite{Planck:2018jri}, we can determine the value of $m$. In Fig. \ref{fR:phi2}, by applying Eq. \eqref{fN:phi2}, we plot the evolution of the function $f(R)$ against the $e$-fold number $N$ by taking $\beta=0.01$ for some typical values of $\gamma$. The prediction of general relativity (GR) for the Ricci scalar $R$ is also shown in the figure by the red color. The figure clearly shows the value of the function $f(R)$ depends on the parameter $\gamma$. As we see, different values of $\gamma$ lead to the different values of $f(R)$. It can be concluded that $\gamma$ manages the behavior of the $f(R)$ function and causes the deviation from GR. Since $\gamma$ is a free parameter in our model, we may take its value such that the regime of the $f(R)$ gravity dominates over the GR regime throughout inflation, and consequently any transition would not occur during inflation at all. This means that in the effective action which may contain the $f(R)$ term together with the Einstein-Hilbert term $R$, the contribution of the latter will be negligible compared to the former contribution, and therefore the effective action will be reduced to the action \eqref{action} which is used in the present work. Besides, since our investigation is not able to determine the precise value of $\gamma$, it is not possible to specify the time of the transition from $f(R)$ to GR that may happen during inflation or in the post-inflationary Universe, in the present work. However, if future studies provide some understanding for us about the nature of the inflaton field, we may determine its effective mass, and accordingly estimate value of the $\gamma$. This enables us to estimate the time of such a transition. In addition, we may provide some observational constraints of our model parameter $\beta$ in the setup of the Tsallis inflation, like the analysis performed in \cite{Luciano:2022pzg} for the case of the Barrow cosmology.

\subsection{Natural inflation }
\label{subsection:natural}

The next model that we consider in our investigation is the natural inflation model in which the inflaton field is presumed to be an axion or pseudo-Nambu-Goldstone boson with a cosine-type periodic potential \cite{Freese:1990rb, adams1993natural}
\begin{equation}
\label{potential:natural}
V(\phi)=\Lambda^{4}\left[1+\cos\left(\frac{\phi}{\phi_{0}}\right)\right] \, ,
\end{equation}
where $\Lambda$ is some non-perturbatively generated scale and $\phi_0$ is the scale that determines the curvature of the potential. Both of these constants have dimensions of mass. It seems that the super-Planckian value of the scale $\phi_0$, i.e. $\phi_0\gtrsim M_P$, is impossible in the context of string theory because all known controlled string theory constructions are restricted to $\phi_0 < M_P$ \cite{svrcek2006axions, banks2003possibility}. In the framework of standard slow-roll inflation, the prediction of the natural potential in the $r-n_{s}$ plane is in tension with the latest observations \cite{Planck:2018jri}, in the sense that its prediction can lie within the 95\% CL region of these data only for $N_*=60$. This issue motivates us to study natural inflation in the Tsallis entropy-based scenario to see whether the novel framework can improve the consistency of the model with the Planck 2018 observations \cite{Planck:2018jri}.

Using Eqs. \eqref{Geff}, \eqref{R:SR}, \eqref{FR1:Tsallis:SR}, and \eqref{fRphi}, the slow-roll parameters in Eqs.  \eqref{epsilon1}, \eqref{epsilon2},  \eqref{epsilon3}, and \eqref{epsilon4} for the Tsallis entropy-based model described by the potential \eqref{potential:natural} can be obtained as
\begin{align}
\label{epsilon1phi:natural}
&\varepsilon_{1}\simeq \frac{3^{\frac{-1+\beta}{2-\beta}}\eta}{2(2-\beta)} \left(\frac{2^{3-2\beta}\pi^{2-\beta}(2-\beta)(1+\cos(\chi))}{\beta}\right)^\frac{1}{\beta-2}(1-\cos(\chi)) \, ,
\\
\label{epsilon2phi:natural}
&\varepsilon_{2}\simeq \frac{3^{\frac{-1+\beta}{2-\beta}}\eta}{2(2-\beta)}
\left(\frac{2^{3-2\beta}\pi^{2-\beta}(2-\beta)(1+\cos(\chi))}{\beta}\right)^\frac{1}{\beta-2}
\Big(1-(2\beta-3)\cos(\chi)\Big) \, ,
\\
\label{epsilon3phi:natural}
&\varepsilon_{3}\simeq \frac{3^{\frac{-1+\beta}{2-\beta}}(-1+\beta)\eta}{2(2-\beta)}
\left(\frac{2^{3-2\beta}\pi^{2-\beta}(2-\beta)(1+\cos(\chi))}{\beta}\right)^\frac{1}{\beta-2}(1-\cos(\chi)) \, ,
\\
\label{epsilon4phi:natural}
&\varepsilon_{4}\simeq \frac{3^{\frac{-1+\beta}{2-\beta}}(-1+\beta)\eta}{(2-\beta)}
\left(\frac{2^{3-2\beta}\pi^{2-\beta}(2-\beta)(1+\cos(\chi))}{\beta}\right)^\frac{1}{\beta-2}(1-\cos(\chi))
\nonumber
\\
&
\times\Bigg\{ (4\pi)^\beta (-1+\beta) \big(1-(\beta-1)\cos(\chi)\big)\eta
+8\pi^2 (-2+\beta)^2(1+\cos(\chi))
\nonumber
\\
&
\times 3^{\frac{1}{\beta-2}}
\left(\frac{2^{3-2\beta}\pi^{2-\beta}(2-\beta)(1+\cos(\chi))}{\beta}\right)^\frac{-\beta+1}{\beta-2}
\Bigg\}
\Big /
\Bigg\{(4\pi)^\beta (-1+\beta)^2 (1-\cos(\chi))\eta
\nonumber
\\
&+ (4\pi)^2 3^{\frac{1}{\beta-2}}(-2+\beta)^2
\left(\frac{2^{3-2\beta}\pi^{2-\beta}(2-\beta)(1+\cos(\chi))}{\beta}\right)^\frac{-\beta+1}{\beta-2}
(1+\cos(\chi))
\Bigg \}.
\end{align}
Here, we have defined
\begin{equation}
\label{eta-chi}
\eta \equiv \frac{\Lambda^4}{{\phi_0}^2}\left(\frac{\Lambda^4}{\gamma}\right)^\frac{1}{\beta-2} \, ,
\hspace{1cm} \chi \equiv \frac{\phi}{\phi_0}.
\end{equation}

Applying Eqs. \eqref{Geff}, \eqref{R:SR}, \eqref{FR1:Tsallis:SR}, \eqref{fRphi}, and \eqref{potential:natural}, we can obtain the scalar power spectrum \eqref{Psphi} as the following form
\begin{align}
\label{Ps:natural}
&{\cal P}_{s}\simeq \Bigg\{2 (-2+\beta)^2 \eta ^{-3} (\Xi)^{4} 3^{\frac{2\beta}{-2+\beta}}
\left(\frac{2^{3-2\beta}\pi^{2-\beta}(2-\beta)(1+\cos(\chi))}{\beta}\right)^\frac{2+\beta}{2-\beta}
\csc^{2}(\frac{\chi}{2})\Bigg\}
\Big/
\nonumber
\\
&
\Bigg\{(4\pi)^\beta (-1+\beta)^2 (1-\cos(\chi))\eta + 16 \pi^{2} (-2+\beta)^2\, 3^{\frac{1}{-2+\beta}} \left(\frac{2^{3-2\beta}\pi^{2-\beta}(2-\beta)(1+\cos(\chi))}{\beta}\right)^\frac{\beta-1}{2-\beta}
\nonumber
\\
&
\times (1+\cos(\chi))\Bigg \} \, ,
\end{align}
where
\begin{equation}\label{xi}
\Xi\equiv \Lambda/\phi_{0} \, .
\end{equation}

If we apply Eq. \eqref{Geff} and also the potential \eqref{potential:natural} in Eq. \eqref{dphitodN:SR}, we reach the following differential equation
\begin{equation}
\label{dphitodN:natural}
\frac{d\chi}{dN}\simeq -3^\frac{-1+\beta}{2-\beta}\eta \left(\frac{2^{3-2\beta}\pi^{2-\beta}(2-\beta)(1+\cos(\chi))}{\beta}\right)^{\frac{1}{\beta-2}}\sin(\chi) \, .
\end{equation}
Solving the differential equation \eqref{dphitodN:natural}, we find the evolution of scalar field $\chi$ with respect to the $e$-fold number $N$ in the slow-roll approximation.

By using Eqs. \eqref{Geff}, \eqref{R:SR}, \eqref{FR1:Tsallis:SR}, \eqref{fRphi}, and \eqref{potential:natural} in Eq. \eqref{nsphi}, we find the scalar spectral index $n_s$ as a function of $\chi$, $\beta$, and $\eta$ as
\begin{align}
& n_{s} \simeq 1 + \frac{3^{\frac{1-\beta}{-2+\beta}}}{2(-2+\beta)}
\Bigg\{5\, \pi^{\beta}\, 2^{1+2\beta}\,3^{1+\beta}
-19\, \beta \, 3^{1+\beta}(4\pi)^{\beta}+2^{3+2\beta}\,3^{1+\beta}\,\pi^{\beta}\beta^{2}
\nonumber\\
& +3^{1+\beta}(4\pi)^{\beta}\beta^{3}
+\eta^{-1}
\Big(448\, \pi^{2}\, 3^{\frac{-1-\beta+\beta^{2}}{-2+\beta}}-64\, \pi^{2}\, 3^{\frac{-5+\beta+\beta^{2}}{-2+\beta}}\beta
+80\, \pi^{2}\, 3^{\frac{-3+\beta^{2}}{-2+\beta}}\beta^{2}
\nonumber\\
&
-32\, \pi^{2}\, 3^{\frac{-1-\beta+\beta^{2}}{-2+\beta}}\beta^{3}\Big)
\left(\frac{2^{3-2\beta}\pi^{2-\beta}(2-\beta)(1+\cos(\chi))}{\beta}\right)^{\frac{\beta-1}{2-\beta}}
\nonumber\\
& -\cos(\chi)\Big[12^{1+\beta}\pi^{\beta}(1-\beta)^{2}(2+\beta)+64\, \pi^{2}\, 3^{\frac{-1-\beta+\beta^{2}}{-2+\beta}}(-2+\beta)^{3}\, \eta^{-1}
\nonumber\\
& \times\left(\frac{2^{3-2\beta}\pi^{2-\beta}(2-\beta)(1+\cos(\chi))}{\beta}\right)^{\frac{\beta-1}{2-\beta}}\Big]
-\cos(2\chi)\Big[-3^{1+\beta}(4\pi)^{\beta}(7+3\beta^{2})\beta
\nonumber\\
&
+2^{1+2\beta}\,3^{1+\beta}\pi^{\beta}(1+4\beta^{2})+16\,\pi^{2}\,3^{\frac{-1-\beta+\beta^{2}}{-2+\beta}}(2-\beta^{2})
(-1+2\beta)\, \eta^{-1}
\nonumber\\
& \times\left(\frac{2^{3-2\beta}\pi^{2-\beta}(2-\beta)(1+\cos(\chi))}{\beta}\right)^{\frac{\beta-1}{2-\beta}}\Big]\Bigg\}
\Big/
\Bigg\{3^{1+\beta}(1-\cos(\chi))
\nonumber\\
& \times(4\pi)^{\beta}(1-\beta)^{2}\eta^{-1}\left(\frac{2^{3-2\beta}\pi^{2-\beta}(2-\beta)(1+\cos(\chi))}{\beta}\right)^{\frac{1}{2-\beta}}
\nonumber\\
& +16\, \pi^{2}(1+\cos(\chi))\, 3^{\frac{-1-\beta+\beta^{2}}{-2+\beta}}(2-\beta)^{2}\eta^{-2}
\left(\frac{2^{3-2\beta}\pi^{2-\beta}(2-\beta)(1+\cos(\chi))}{\beta}\right)^{\frac{\beta}{2-\beta}}\Bigg\} \, .
\label{ns:natural}
\end{align}

We also can rewrite the tensor-to-scalar ratio $r$ in Eq. \eqref{rphi} as the following form
\begin{align}
\label{r:natural}
r\simeq & \frac{3^{\frac{2-\beta^{2}}{-2+\beta}}\,\beta\,\eta}{(\beta-2)^{3}\, \pi^{2}}
\left(\frac{2^{3-2\beta}\, \pi^{2-\beta}(2-\beta)}{\beta}\right)^{\frac{1+\beta}{\beta-2}}
\cos^{2}\big(\frac{\chi}{2}\big)\big(1+\cos(\chi)\big)^{\frac{2(\beta-3)}{2-\beta}}
\nonumber
\\
&
\times\Big[3^{1+\beta}(4\pi)^{\beta}(1-\beta)^{2}\,\eta
\left(\frac{2^{3-2\beta}\,\pi^{2-\beta}(2-\beta)}{\beta}\right)^{\frac{1}{2-\beta}}\big(\cos(\chi)-1\big)
\nonumber
\\
&
-16\, \pi^2 (2-\beta)^2 \, 3^\frac{-1-\beta+\beta^2}{-2+\beta}\big(1+\cos(\chi)\big)^\frac{1}{2-\beta}
\left(\frac{2^{3-2\beta}\,\pi^{2-\beta}(2-\beta)}{\beta}\right)^\frac{\beta}{2-\beta}
\Big] \sin^2(\chi) \, ,
\end{align}
where we have applied Eqs. \eqref{Geff}, \eqref{R:SR}, \eqref{FR1:Tsallis:SR}, \eqref{fRphi}, and \eqref{potential:natural} in deriving Eq. \eqref{r:natural}.

With the help of Eqs. \eqref{Geff} and \eqref{potential:natural}, we can rewrite Eq. \eqref{fRphi} as
\begin{equation}\label{fphi:natural}
f=\frac{4\Lambda^4}{\beta}(1+\cos(\chi)) \, .
\end{equation}

Substituting Eqs. \eqref{Geff} and \eqref{potential:natural} in Eq. \eqref{FR1:Tsallis:SR}, the Hubble parameter $H$
is found as the following form
\begin{equation}\label{Hphi:natural}
H=3^\frac{1}{2(\beta-2)}\,\Big(\frac{\Lambda^4}{\phi_0^{2}\eta}\Big)^\frac{1}{2}
\left(\frac{2^{3-2\beta}\,\pi^{2-\beta}(2-\beta)(1+\cos(\chi))}{\beta}\right)^\frac{1}{2(2-\beta)} \, .
\end{equation}
We can apply the above equation together with Eq. \eqref{aN} in Eq. \eqref{kdefin}, to find the comoving number $k$ as
\begin{equation}\label{kN:natural}
k(N)=\Big((1+\cos(\chi))^\frac{1}{2(2-\beta)}(1+\cos(\chi_*))^\frac{1}{2(-2+\beta)}\,\exp({N_{*}-N})\Big)\,
0.05 \, {\rm Mpc}^{{\rm -1}} \, .
\end{equation}

To find the scalar field $\chi$ in terms of the $e$-fold number $N$, we perform the same steps as the previous subsection. In the first step, using Eq. \eqref{epsilon1phi:natural} for a given set of the model parameters $(\beta, \eta)$, we solve the equation $\varepsilon_{1}(\chi_{e})=1$ numerically to find $\chi_{\rm e}$. After that, with the help of the obtained value of $\chi_{\rm e}$, we solve differential equation \eqref{dphitodN:natural}, numerically, and consequently we find $\chi=\chi(N)$. Substituting $\chi(N)$ into Eqs. \eqref{epsilon1phi:natural}, \eqref{epsilon2phi:natural}, \eqref{epsilon3phi:natural}, \eqref{epsilon4phi:natural}, \eqref{ns:natural}, \eqref{r:natural}, and \eqref{kN:natural}, the slow-roll parameters $\varepsilon_{1}$, $\varepsilon_{2}$, $\varepsilon_{3}$, $\varepsilon_{4}$, the inflationary observables $n_s$ and $r$, and also the comoving number $k$ are derived in terms of $N$, respectively.

\begin{figure}[t]
\begin{center}
\scalebox{0.9}[0.9]{\includegraphics{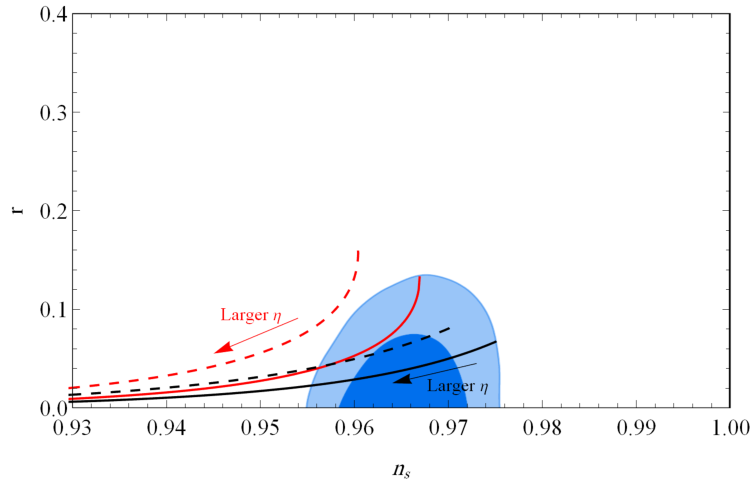}}
\caption{The $r-n_s$ diagram of the natural potential \eqref{potential:natural} in the Tsallis inflationary setting by taking $\beta=1/2$ for two different values of $N_*$ with varying $\eta$ in the range of $\eta>0$. The results of our model for $N_*=50$ and $N_*=60$ are shown by the dashed and solid black curves, respectively.
The red dashed ($N_*=50$) and solid ($N_*=60$) curves show the prediction of the standard natural inflation. The marginalized joint 68\% and 95\% CL regions of the Planck 2018 TTTEEE+low$\ell$+lowE data \cite{Planck:2018jri} are specified by dark and light blue, respectively.}
\label{rns:natural}
\end{center}
\end{figure}

In Fig. \ref{rns:natural}, we plot the $r-n_{s}$ diagram for $\beta= 1/2$ and $N_*=50, 60$ with varying $\eta$ in the range of $\eta>0$. The black dashed and solid curves illustrate the prediction of the model for $N_*=50$ and $N_*=60$, respectively. The standard natural inflation which corresponds to $\beta=1$ in our model, is also presented by the red dashed ($N_*=50$) and solid ($N_*=60$) curves. From the figure, it is obvious that as $\eta$ increases, both $n_s$ and $r$ get smaller. Moreover, Fig. \ref{rns:natural} clears that the result of the potential \eqref{potential:natural} in the standard inflation for $N_*=50$ lies completely outside the allowed regions of the Planck 2018 observations \cite{Planck:2018jri} and only for $N_*=60$ enters the 95\% CL region of these data. In the Tsallis entropy-based inflationary setting, however, its prediction can lie well within the 68\% CL region allowed by the Planck 2018 data \cite{Planck:2018jri}.

Using the Planck observational constraints on $r–n_{s}$ plane and also Eqs. \eqref{ns:natural} and \eqref{r:natural}, we also can estimate the ranges of parameter $\eta$ for which the prediction of our model for $\beta=1/2$ and different values of $N_*$ is compatible with 68\% CL constraint of the Planck 2018 data \cite{Planck:2018jri}. In the case $N_*=50$, the model is compatible with 68\% CL constraint of the Planck 2018 data for $0.172\lesssim \eta \lesssim 0.433$. We also find that for $N_*=60$, the results of our model can lie inside the 68\% CL region of the Planck 2018 data, if $0.281 \lesssim \eta \lesssim 0.613$.

To present some constraint on the model parameter $\eta$ by using the recent constraint of \cite{Galloni:2022mok} on $r_{0.01}$, we determine the $e$-fold number at which the comoving wavenumber $k=0.01\,\mathrm{Mpc}^{-1}$ exits the Hubble horizon during inflation, with the help of Eq. \eqref{kN:natural} by taking $\beta=1/2$ and $N_{*}=60$. It is found that $N_{0.01} \simeq 61.6$.

\begin{figure}[t]
\begin{center}
\scalebox{0.9}[0.9]{\includegraphics{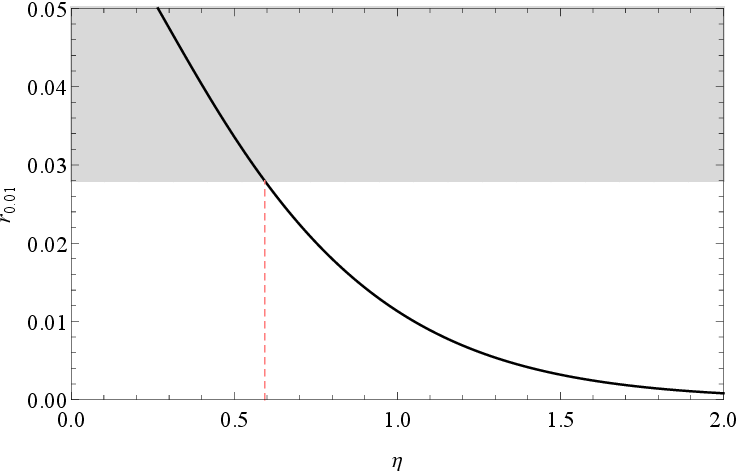}}
\caption{Variation of the tensor-to-scalar ratio $r_{0.01}$ against the parameter $\eta$ for the natural potential \eqref{potential:natural} in the Tsallis inflationary scenario with $\beta=1/2$ and $N_{0.01} \simeq 61.6$. The gray-shaded region is excluded by the constraint on the upper bound on $r_{0.01}$, reported by Galloni et al. \cite{Galloni:2022mok}. The pink vertical dashed line specifies $\eta \simeq 0.593$ which is the minimum value of the parameter $\eta$ for which the model agrees with the constraint $r_{0.01}<0.028$.}
\label{reta:natrual}
\end{center}
\end{figure}

In Fig. \ref{reta:natrual}, using Eq. \eqref{r:natural}, we plot the variation of the tensor-to-scalar ratio $r_{0.01}$ against the parameter $\eta$ by setting $\beta=1/2$ and $N_{0.01} \simeq 61.6$. The gray-shaded region is excluded by the constraint on the upper bound on $r_{0.01}$ reported by Galloni et al. \cite{Galloni:2022mok}. The prediction of the model is in agreement with this constraint if $\eta \gtrsim 0.593$.

The evolution of the slow-roll parameters $\varepsilon_{1}$, $\varepsilon_{2}$, $\varepsilon_{3}$, and $\varepsilon_{4}$ versus the $e$-fold number $N$ for the natural potential \eqref{potential:natural} is plotted in Fig. \ref{SRparameters:phi2}, by using Eqs. \eqref{epsilon1phi:natural}-\eqref{epsilon4phi:natural}. In this figure, we have considered $\beta=1/2$ and $\eta=0.65$. The figure shows that the slow-roll approximation holds during inflation. This point verifies the viability of our investigation which is based on the slow-roll approximation. Moreover, from the figure, we see that the first slow-roll parameter $\varepsilon_{1}$ reaches unity at the end of inflation with $N = 0$.

\begin{figure}[t]
\begin{center}
\scalebox{0.9}[0.9]{\includegraphics{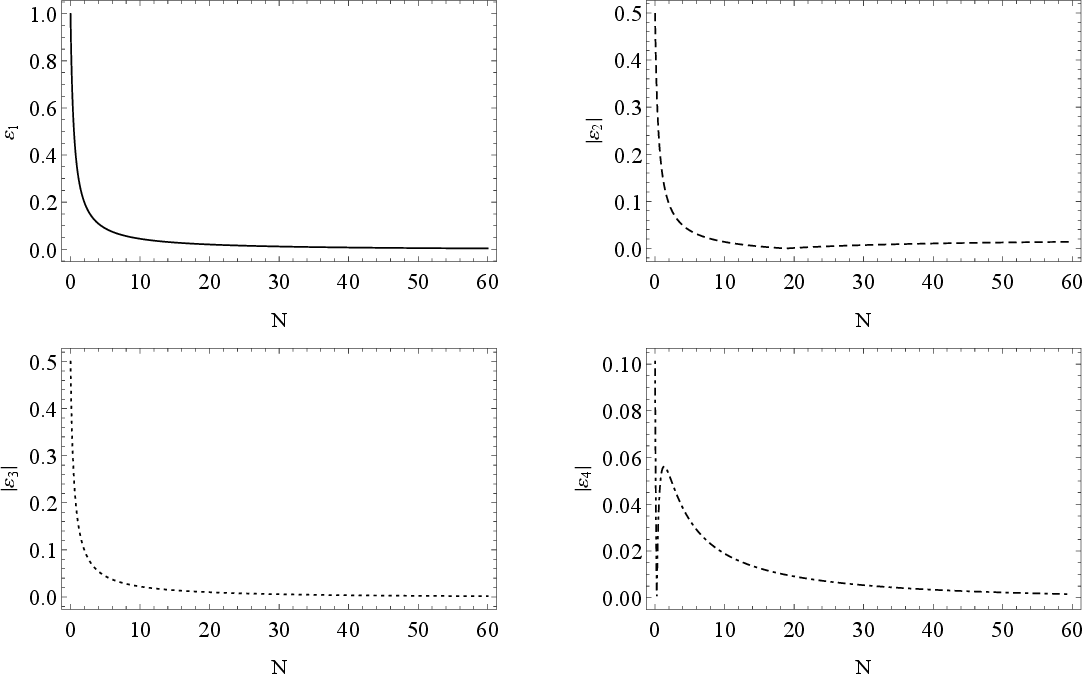}}
\caption{Evolution of the slow-roll parameters $\varepsilon_{1}$, $\varepsilon_{2}$, $\varepsilon_{3}$, and $\varepsilon_{4}$ as a function of the $e$-fold number $N$ for the natural potential \eqref{potential:natural} in the Tsallis entropy-based model
with $\beta=1/2$ and $\eta=0.65$.}
\label{SRparameters:phi2}
\end{center}
\end{figure}

\begin{figure}[t]
\begin{center}
\scalebox{0.9}[0.9]{\includegraphics{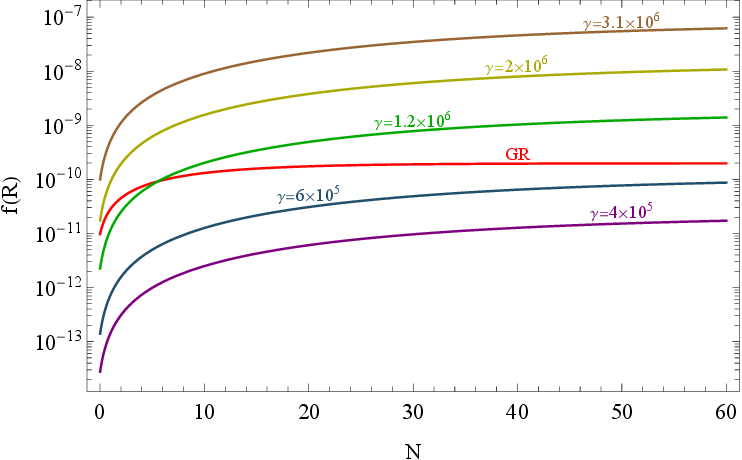}}
\caption{Evolution of the function $f(R)$ against of the $e$-fold number $N$ for the natural potential \eqref{potential:natural} in the Tsallis inflation with $\beta=1/2$ and $\eta=0.65$, for some typical values of $\gamma$. The prediction of general relativity (GR) for the Ricci scalar $R$ is also shown by the red color.}
\label{fR:natrual}
\end{center}
\end{figure}

In Fig. \ref{fR:natrual}, with the help of Eq. \eqref{fphi:natural}, the evolution of the function $f(R)$ as a function of the $e$-fold number $N$ is plotted for some typical values of $\gamma$, and by setting $\beta=1/2$ and $\eta=0.65$. Note that to find the parameter $\Lambda$, we follow this approach: in the first step, for given values of the parameters $\beta$ and $\eta$, the value of $\Xi$ is fixed by imposing the CMB normalization at the pivot scale $k_{*}=0.05\, {\rm Mpc}^{{\rm -1}}$ corresponding to $N_{*}=60$, with the help of Eq. \eqref{Ps:natural}. Afterwards, for a specific value of the parameter $ \gamma$ by using the first definition in Eq. \eqref{eta-chi} and also Eq. \eqref{xi}, one can find the value of the parameter $\Lambda$. In the figure, We also show the Ricci scalar $R$ predicted by GR by the red color. Fig. \ref{fR:natrual} clears the effect of the parameter $\gamma$ on the function $f(R)$. Since in this model we are free to take any values for the parameter $\gamma$, by choosing an appropriate value of this parameter, we can guarantee that the regime of $f(R)$ gravity dominates over the GR regime in the whole of inflation. Therefore, if the effective action includes the Einstein-Hilbert term $R$ in addition to the $f(R)$ term, then the former contribution can be dropped versus the latter contribution, and the effective action turns into the action \eqref{action}. To determine the epoch of the transition from $f(R)$ to GR, we need the precise value of the $\gamma$ parameter that cannot be determined in the present work.

\section{Conclusions}
\label{section:conclusions}

Inflation has occurred in the regime of high energy physics at which the gravitational theory is expected to be modified. Therefore, the entropy-area relation may undergo some modifications in those energy scales. This motivated us to regard the entropy of the early Universe to be in the form of the Tsallis entropy which is a generalization for the Bekenstein-Hawking entropy \cite{bekenstein1973black} and possesses the non-additivity and non-extensivity property. This form of entropy has a relation with the horizon area as $S_h=\gamma A^{\beta}$, in which $A$ is the area of the horizon and $\beta$ and $\gamma$ are unknown constants. We have studied the inflationary era in the context of the Tsallis entropy-based cosmology. Since there is no definite action for this setup, it was not possible to derive the power spectra of the primordial scalar and tensor perturbations in our Tsallis inflation scenario. To resolve this issue, we reconstructed an $f(R)$ model which is thermodynamically equivalent to our setting in the slow-roll approximation. This equivalence allows us to use the equations of the scalar and tensor power spectra obtained in the $f(R)$ gravity for our Tsallis inflation model. We have considered two different inflationary potentials in our scenario and checked their viability with the Planck 2018 observations \cite{Planck:2018jri}.

First, we studied the observational consistency of the quadratic potential \eqref{potential:phi2}  which provides a chaotic inflation model. In the standard inflationary setting based on the Einstein gravity, the consistency of this potential is not favored by the Planck 2018 observational data \cite{Planck:2018jri}. This motivated us to investigate whether this potential can be resurrected in light of the Planck 2018 results in the setting of the Tsallis entropy-based inflation. We have derived analytic formulas for $n_s$ and $r$ in terms of the parameter $\beta$ and the $e$-fold number $N$, and then plotted the $r-n_{s}$ diagram for $N_*=50, 60$, with varying the parameter $\beta$ in the range of $0<\beta<2$. Our results imply that this potential can be in excellent consistency with the Planck 2018 data in the Tsallis entropy-based scenario, such that its results can lie inside the 68\% CL region of the observational data.

Moreover, we have proved that in the limits $\beta \ll 1$ and $N\gg1$, the behavior of the quadratic potential in the $r-n_s$ plane coincides exactly with the prediction of the Starobinsky $R^2$ inflation \cite{Starobinsky:1980te}. Furthermore, we have estimated that for $N_*=50$, the prediction of the model is compatible with the 68\% CL constraint of the Planck 2018 observations, if $0< \beta\lesssim 0.045$. In the case of $N_*=60$, the model is consistent with the 68\% CL region of the observational data provided that $ 0<\beta\lesssim 0.011$. The recent constraint $r_{0.01}<0.028$ (95\% CL) \cite{Galloni:2022mok} on the tensor-to-scalar ratio at the scale $k=0.01\,\mathrm{Mpc}^{-1}$ constrains this parameter to $0<\beta\lesssim 0.202$.

We further examined the viability of the natural potential \eqref{potential:natural} in the inflationary setting based on the Tsallis entropy. In the framework of the standard inflation, the prediction of this potential in the $r-n_{s}$ plane is not very preferred according to the current CMB observations, regarding the fact that its results can satisfy only the 95\% CL constraints of the Planck 2018 data \cite{Planck:2018jri}. In this case, we have found the inflationary observables $n_s$ and $r$ in terms of the model parameters $\beta$ and $\eta$ numerically. We have focused on the cases $\beta=1/2$ and drawn the $r-n_{s}$ diagram for $N_*=50, 60$, with varying $\eta$ in the range of $\eta>0$. We have demonstrated that in the framework of the Tsallis entropy-based inflation, the natural potential provides a great fit to the Planck 2018 data \cite{Planck:2018jri}. The prediction of the model lies inside the 68\% CL region of these data. Furthermore, we presented some observational constraints on the model parameters by using the Planck 2018 data \cite{Planck:2018jri}. Our results imply that for $N_*=50, 60$, the results of our model can lie inside the 68\% CL region of the Planck 2018 data \cite{Planck:2018jri}, provided that $0.172\lesssim \eta \lesssim 0.433$ and $0.281 \lesssim \eta \lesssim 0.613$, respectively. The observational bound on $r_{0.01}$ provided in \cite{Galloni:2022mok}, also gives rise to the condition $\eta\gtrsim 0.593$ for the model parameter.


 \begin{acknowledgments}
The authors thank the referees for their valuable comments.
\end{acknowledgments}






%




\end{document}